\documentclass[aps,prl,reprint,superscriptaddress]{revtex4-2}
\usepackage[utf8]{inputenc}
\usepackage{hyperref}
\hypersetup{
	unicode,
	pdfauthor={Reinier van der Meer},
	pdftitle={Semi-device-independent indistinguishability witness},
	pdfsubject={Semi-device-independent indistinguishability witness},
	pdfkeywords={Witness, Quantum interference, Certification, Multi-photon interference},
	pdfproducer={LaTeX},
	pdfcreator={pdflatex}
}

\usepackage{graphicx}
\graphicspath{{Figures/}}

\usepackage{amsmath}
\usepackage{amsthm}
\usepackage{amssymb}
\usepackage{times}
\usepackage{braket}

\usepackage{ulem}
\usepackage{color}

\begin{document}

\title{Experimental demonstration of an efficient, semi-device-independent photonic indistinguishability witness}

\author{Reinier van der Meer}
\author{Peter Hooijschuur}
\author{Franciscus H.B. Somhorst}
\affiliation{Adaptive Quantum Optics group, University of Twente, PO Box 217, 7500 AE Enschede, The Netherlands.}

\author{Pim Venderbosch}
\author{Michiel de Goede}
\author{Ben Kassenberg}
\author{Henk Snijders}
\author{Caterina Taballione}
\author{J\"{o}rn Epping}
\author{Hans van den Vlekkert}
\affiliation{Quix Quantum BV, Hengelosestraat 500, 7521 AN Enschede, The Netherlands}

\author{Nathan Walk}
\affiliation{Dahlem Center for Complex Quantum Systems, Freie Universit{\"a}t Berlin, 14195 Berlin, Germany}

\author{Pepijn W.H. Pinkse}
\affiliation{Adaptive Quantum Optics group,
University of Twente, PO Box 217, 7500 AE Enschede, The Netherlands.}

\author{Jelmer J. Renema}
\email[]{j.j.renema@utwente.nl}
\affiliation{Adaptive Quantum Optics group,
University of Twente, PO Box 217, 7500 AE Enschede, The Netherlands.}
\affiliation{Quix Quantum BV, Hengelosestraat 500, 7521 AN Enschede, The Netherlands}

\date{\today}

\begin{abstract}
    Efficient and reliable measurements of photonic indistinguishability are crucial to solidify claims of a quantum advantage in photonics. Existing indistinguishability witnesses may be vulnerable to implementation loopholes, showing the need for a measurement which depends on as few assumptions as possible. Here, we introduce a semi-device-independent witness of photonic indistinguishability and measure it on an integrated photonic processor, certifying  three-photon indistinguishability in a way that is insensitive to implementation errors in our processor.
\end{abstract}

\maketitle

\section{Introduction}
Quantum photonics has shown substantial potential as a platform for near-term quantum information processing, culminating in the recent demonstrations of a quantum advantage in boson sampling \cite{zhong_2020_Science,zhong_2021_ArXiv}, i.e., of experiments in which a quantum device outperforms a classical one at a certain computational task. In these experiments \cite{zhong_2020_Science,zhong_2021_ArXiv,broome_2013_Science,spring_2013_Science,tillmann_2013_NatPhotonics,crespi_2013_NatPhotonics,wang_2017_Nat.Photonics,He2017,Paesani2019,Zhong2020,Zhong2021,Arrazola2021}, photons are sent through a large-scale linear interferometer, which can be implemented either in static form (typically in bulk optics) or as a programmable interferometer, typically in integrated optics. An advantage arises because it is believed to be classically computationally hard to draw samples from the probability distribution that arises when the output state of the interferometer is measured in the Fock basis \cite{aaronson_2013_TheoryComputa,deshpande_2021_ArXiv}.

These experiments operate at the edge of what is technologically possible, requiring almost perfect optical systems. Any noise introduced by an imperfect optical system may open up a loophole, making the system susceptible to efficient classical simulation \cite{renema_2018_Phys.Rev.Lett.,shchesnovich_2020_Int.J.QuantumInform.,villalonga_2021_ArXiv}. In particular, it is required that the optical system has low loss \cite{aaronson_2016_PhysRevAa,renema_2018_ArXiv,garcia-patron_2019_Quantum,Brod2020}, and that the photons have a high degree of indistinguishability \cite{renema_2018_Phys.Rev.Lett.}, i.e. that the photons are identical in those aspects of their wave function not perturbed by the interferometer, such as wavelength, polarization, and temporal structure.

Indistinguishability is central in linear-optical information processing, both in the boson sampling paradigm and for universal linear-optical quantum computation \cite{knill_2001_Nature,browne_2005_Phys.Rev.Lett.,aaronson_2013_TheoryComputa}, because it enables the creation of entanglement between non-interacting particles. This entanglement arises because a quantum state comprised of multiple indistinguishable bosons must have an overall symmetric wave function. The Hong-Ou-Mandel effect \cite{Hong1987} is the archetypical example of this process, and the quantum advantage experiments of \cite{zhong_2020_Science,zhong_2021_ArXiv} can be thought of as multiphoton, multimode generalizations of this effect. Accurate measurements of the photon indistinguishability \cite{tichy_2014_Phys.Rev.Lett.,carolan_2014_NaturePhoton,bentivegna_2014_Int.J.QuantumInform.,crespi_2015_Phys.Rev.A,bentivegna_2016_NewJ.Phys.,Walschaers2016,dittel_2017_QuantumSci.Technol.,viggianiello_2018_ScienceBulletin,giordani_2018_NaturePhoton,viggianiello_2018_NewJ.Phys.,agresti_2019_Phys.Rev.X,brod_2019_Phys.Rev.Lett.} are therefore a central question in the discussion surrounding the classical simulability of recent quantum advantage experiments. These measurements must be done in an experimental configuration as close to that in which a quantum advantage is obtained, and ideally on all pairs of photons simultaneously. 

To address this question, the notion of an indistinguishability witness was introduced \cite{brod_2019_Phys.Rev.Lett.}. Analogous to entanglement witnesses \cite{Terhal2000}, for a given linear network $U$, an indistinguishability witness is an observable $\hat{C}(U)$ whose expected value $C(U)$ has some upper bound $C_c(U)$ for distinguishable particles, but whose value for indistinguishable particles $C_q(U)$ exceeds that bound. When we measure $C$ in an experiment and find a value which exceeds $C_c(U)$, we can therefore infer that the photons used are indistinguishable. 

However, such witnesses are open to implementation loopholes \cite{Shchesnovich2016}: given witness $C(U)$, it is possible that there will be some other linear network $U'$ which has a larger value of $C_{c}(U') > C_{q}(U)$. If the linear interferometer is then set to $U'$, indistinguishability would be falsely certified. Given the fact that control of programmable optics is not perfect, this is a problem of concern, being analogous to the situation in entanglement witnesses, which can be compromised if an adversarial party takes over a trusted detection apparatus \cite{Gerhardt2011}. 

For the analogous problem of certifying other quantum properties such as entanglement or randomness, an array of approaches have appeared which relax the assumptions made on the measurement apparatus \cite{BenchmarkingReview}. The most dramatic of these are fully device-independent protocols \cite{pironio2010random,Acin:2016ke}, which rely on Bell nonlocality \cite{Brunner:2014vw} to provide a highly rigorous level of certification but which are extremely experimentally challenging \cite{Liu:2021ba,Shalm:2021jq}. The plethora of intermediate methods that trade practicality for additional assumptions and characterisation are known as \textit{semi-device independent} \cite{pawlowski2011semi,cao2016source,marangon2017source,Drahi:2020gw,chaturvedi2015measurement,cao2015loss,Himbeeck2017semidevice,brask2017megahertz,mattar2017experimental,Kocsis:2015gu}.

In this work, we present a semi-device independent indistinguishability witness, by finding the largest achievable value of $C$  over all linear networks, and choosing that as the threshold value. If the matrix settings are corrupted, the experiment will simply fail to certify indistinguishability, but it will never incorrectly certify a set of distinguishable photons as being indistinguishable. We show that this witness, which is based on two-mode correlators, can be efficiently measured in an experiment by using a large-scale programmable interferometer. We implement this witness in an experiment using a programmable integrated photonic interferometer \cite{Carolan2015, Taballione2019, Taballione2021} and a multiphoton source, and use it to certify the indistinguishability of three photons. Furthermore, we show that the degree to which this witness violates its classical threshold can be used to lower bound the degree of indistinguishability between the photons. We have therefore introduced an efficient method of unambiguously certifying multiphoton indistinguishability, which is directly applicable in large-scale quantum interference experiments.

\section{Indistinguishability witness}

\begin{figure*}[t!b]
    \centering
    \includegraphics{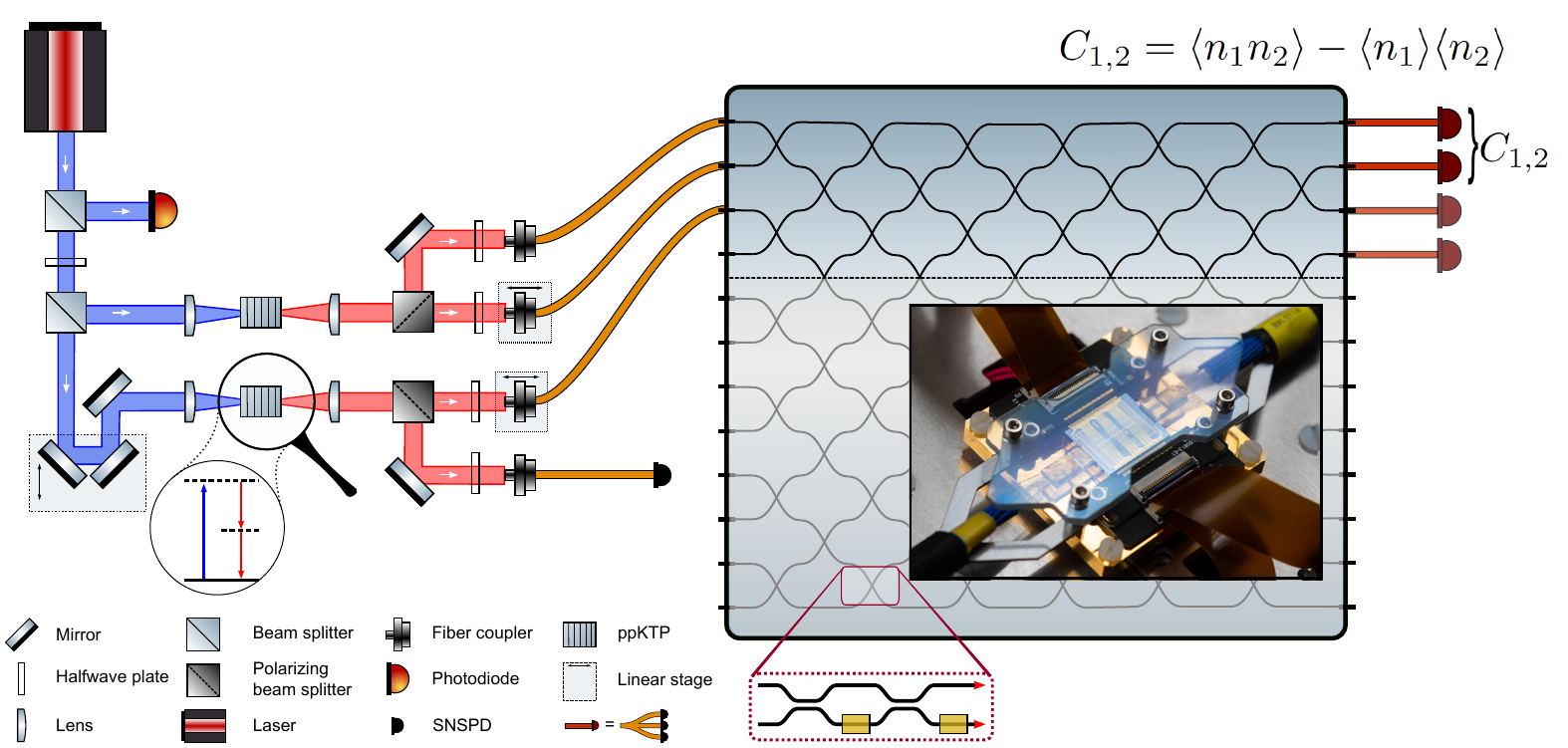}%
    \caption{A $775\,$nm pulsed laser (blue) pumps the two periodically poled KPT (ppKTP) spontaneous parametric down-conversion sources. Each crystal probabilistically generates a pair of two photons with orthogonal polarization (red). One of these four photons is used as a herald and the remaining three are injected in the first three modes of our $12\times12$ integrated universal programmable processor. The processor output is sent to a small fiber beam splitter and SNSPDs acting as pseudo-number-counting detectors. In the processor, the top 4 modes are used to implement the transmission matrices. The zoom-in shows a unit cell of a Mach-Zehnder interferometer that implements one of the programmable beam splitters. The inset shows a photograph of the fiber-connected integrated optical chip used for this experiment.}
    \label{fig:figSetup}
\end{figure*}

The basis of our method are the Ursell or truncated correlation functions \cite{Ursell1927}. Walschaers et al. \cite{Walschaers2016} introduced these functions as $n$-mode correlator functions in a quantum-optical context, to measure the excess correlation among a subset of the optical modes of the interferometer. The two-mode correlation functions were used to perform a statistical certification of photon indistinguishability \cite{Walschaers2016}. We will show that that the extremal values of these correlators form an indistinguishability witness. 
The two-mode correlators are given by: 
\begin{equation}
    C_{i,j} = \braket{\hat{n}_i,\hat{n}_j} - \braket{\hat{n}_i}\braket{\hat{n}_j},
    \label{eq:eqC}
\end{equation}
where $\braket{.}$ denotes the expected value, and $\hat{n}_i$ is the photon number operator acting on optical mode $i$. Considering the case of a linear optical circuit given by a transformation matrix $U$, with one photon injected at each of the first $n$ input ports, which is the standard configuration for boson sampling, this reads:
\begin{equation}
    \begin{aligned}
        C_{i,j} = &-\sum_{k=1}^n |U_{i,k}|^2 |U_{j,k}|^2\\
        &+ \sum_{k\neq l=1}^n x_{kl}^2 U_{i,k} U_{j,l} U_{i,l}^* U_{j,k}^*,\\
    \end{aligned}
    \label{eq:eqCfull}
\end{equation}
where $U$ is the transfer matrix of the linear optical system, and $x_{kl} = |\langle \psi_k|\psi_l \rangle|$ is a quantity measuring the degree of indistinguishability between the $k$-th and $l$-th photons in terms of their internal wave functions, with $x=0$ corresponding to distinguishable photons and $x=1$ corresponding to completely indistinguishable ones. Note that the first term in eq. \ref{eq:eqCfull} corresponds to classical transmission of light, since it is not affected by the degree of distinguishability, whereas the second term describes the quantum interference. Without loss of generality, we restrict ourselves to the first two modes in the network, hence we drop these indices in the notation. 

The correlator can function as a semi-device-independent distinguishability witness if it can lower bound the degree of indistinguishability, \textit{regardless} of the implemented network. We must therefore maximize eq. 2 for distinguishable photons, i.e., when the second term is zero. Any matrix  which obeys the condition $\forall k \leq n, (U_{i,k} = 0) \vee (U_{j,k} = 0)$ maximizes $C$ for distinguishable particles at $C_c = 0$. An example of such a matrix is the identity matrix. The fact that all terms in the first sum in eq. 2 are negative implies that this solution is optimal. This proves that regardless of the matrix implemented, $C_c(U) \leq 0$. This result has a simple interpretation: in the absence of entanglement, the only correlation allowed between particles is via the total photon number, which is conserved. Therefore, if a particle is observed in a particular mode, it decreases the probability of observing a particle in another mode. 

Next, we must find a matrix which has $C_q > 0$. In principle, any matrix which obeys this condition suffices. However, we believe we have found the optimal matrix: we observe that the second term in eq. 2 is maximized if all elements are in phase, and symmetry dictates that the magnitude of all matrix elements should be identical. Our conjectured solution, which is supported by numerical evidence \footnote{See Supplemental Material}, is a matrix of size $N = n+1$, with a highly symmetric form:
\begin{equation}
    \label{eq:optimummatrixrows}
    U_{max} = \begin{pmatrix}
    \frac{1}{\sqrt{2n}} & \cdots & \frac{1}{\sqrt{2n}} & \frac{1}{\sqrt{2}}\\
    \frac{1}{\sqrt{2n}} & \cdots & \frac{1}{\sqrt{2n}} & -\frac{1}{\sqrt{2}}\\
    * & * & * & *\\
    * & * & * & *\\
    \end{pmatrix},
\end{equation}
where $*$ denotes elements that do not enter in the correlator. Note also that since we are only interested in the correlator between the first and second mode, we need only specify the first two rows of the matrix; if these are orthogonal, which is enforced by the $(n+1)$-th column, the Gram-Schmidt process guarantees that we can always complete the matrix in an orthogonal fashion. 

The correlator for this matrix as a function of the indistinguishability follows from eq. \ref{eq:optimummatrixrows}:
\begin{equation}
    C(x) = \frac{-n + \sum_{i,j,i\neq j}x^2_{i,j}}{4n^2}.
    \label{eq:eqCmax}
\end{equation}

To understand the behaviour of this equation, it is instructive to look at the case where $x_{i,j} = 1$, i.e. where all photons are perfectly indistinguishable. In that case, the threshold is $C_q = \frac{1}{4} - \frac{1}{2n}$. 
which is larger than $C_c = 0$ for all $n$ greater than 2; therefore, using this method, we can certify indistinguishability between 3 or more photons. Similarly, if we set all $x_{i,j} = \bar{x}$, then we see that for $n = 3$ photons, no indistinguishability can be certified if $\bar{x}^2 \leq 0.5$. 

To see that measuring this witness is efficient in the number of samples required, observe that the probability that the average photon flux on either of the two modes of interest is $\frac{1}{2}$, irrespective of the number of photons. This means that the probability that a given run of the experiment contributes to the measurement statistics of $C$ is $\frac{3}{4}$. This quantity is independent of $n$ and $N$. Finally, we note that the precision required to certify indistinguishability between $n$ and $n+1$ photons grows quadratically with $n$. These facts together prove that the scheme is efficient in the number of measurements required. 
\section{Setup}
To demonstrate our semi-device-independent indistinguishability witness, we carry out an experiment using a programmable linear optical interferometer, a photon source and a bank of single-photon detectors.
Fig. \ref{fig:figSetup} shows our experimental setup. The optical network is a $12$-mode quantum photonic processor \cite{Taballione2021} implemented in TriPleX silicon nitride waveguides \cite{Roeloffzen2018}. The chip consists of a grid of unit cells, as shown in Fig. \ref{fig:figSetup} \cite{Clements2016}. Each cell consists of a Mach-Zehnder interferometer with two $50:50$ static beam splitters and two tunable phase shifters. Together, these elements can implement any arbitrary pairwise mode interaction. The network is fully programmable using the thermo-optical effect. The total optical transmission through the chip varies between $54$\% to $60$\%, depending on the particular combination of input and output modes. In our experiment, only the top four modes of the chip are used.

The photon source consists of two $2\,$mm ppKTP crystals in a Type-II degenerate configuration. The crystals are pumped with $60\,$mW of $120\,$fs pulsed light at $775\,$nm. Each crystal generates a pair of photons probabilistically at $1550\,$nm \cite{Evans2010}. We herald one photon and postselect on fourfold coincidences to herald the input state $\ket{\psi_{\rm in}} = \ket{1110}$. The photons are $93$\% indistinguishable when they originate from the same crystal (signal-idler), and $78$\% indistinguishable when they are from different crystals (signal-signal). This corresponds to an average three-photon indistinguishability of $x^2 = 0.81$. The average fourfold coincidence rate is $250\,$Hz without the chip.

The photons are detected using $12$ superconducting single-photon detectors \cite{Marsili2011,Holzman2019}. Each output mode of the chip directs the photons towards three detectors, which effectively function as a pseudo-photon-number-resolving detector \cite{Feito2009}. The relative detection efficiencies and splitting ratios are calibrated by sequentially sending photons from input mode 1 to all 4 output modes and comparing the relative count rates. The detectors are read out using standard correlation electronics.

\section{Results}
\begin{figure*}[htb]
    \centering
    \includegraphics[width=1.0\textwidth,keepaspectratio]{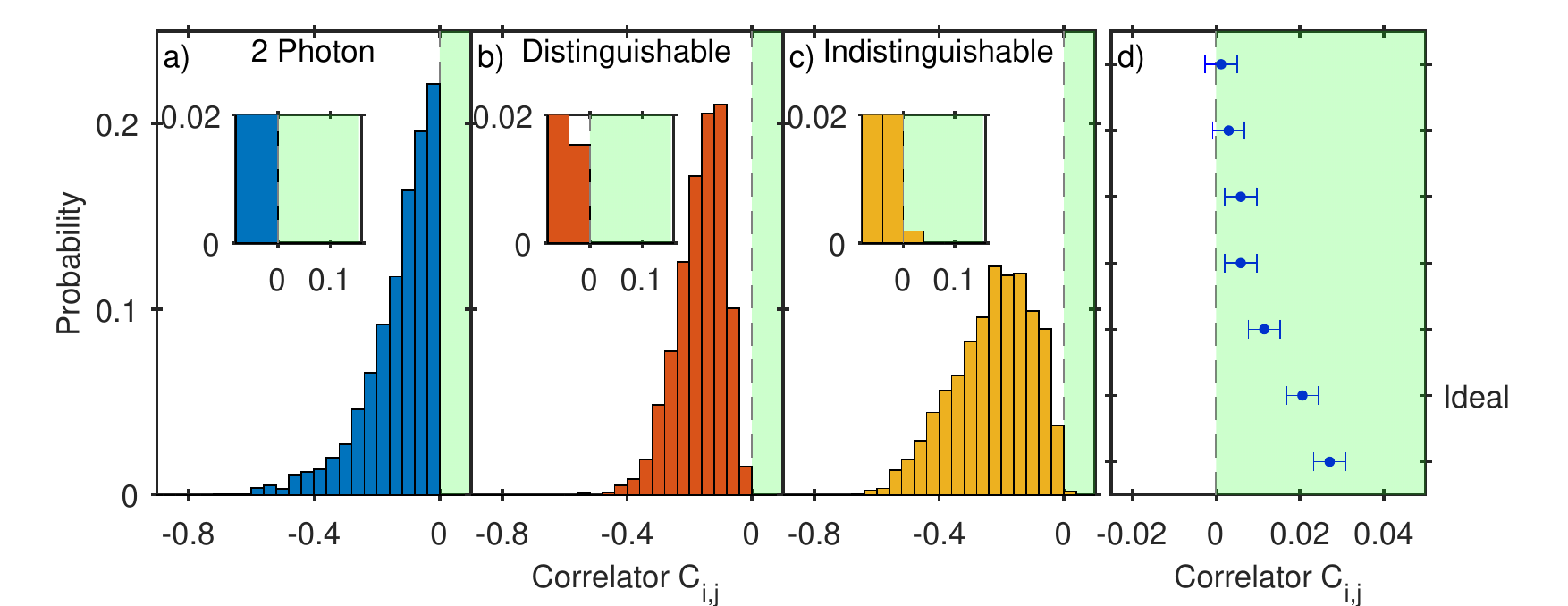}%
    \caption{The distribution of the two-point correlator is never larger than 0 for a) two identical photons and b) three distinguishable photons. For three indistinguishable photons (c), several instances of positive correlators are observed. The insets show a zoom-in at $C=0$. All positive correlators and the corresponding error bar (1 standard deviation) are shown in an ordered fashion in d).}
    \label{fig:figResults}
\end{figure*}

\begin{figure}[tb]
    \centering
    \includegraphics{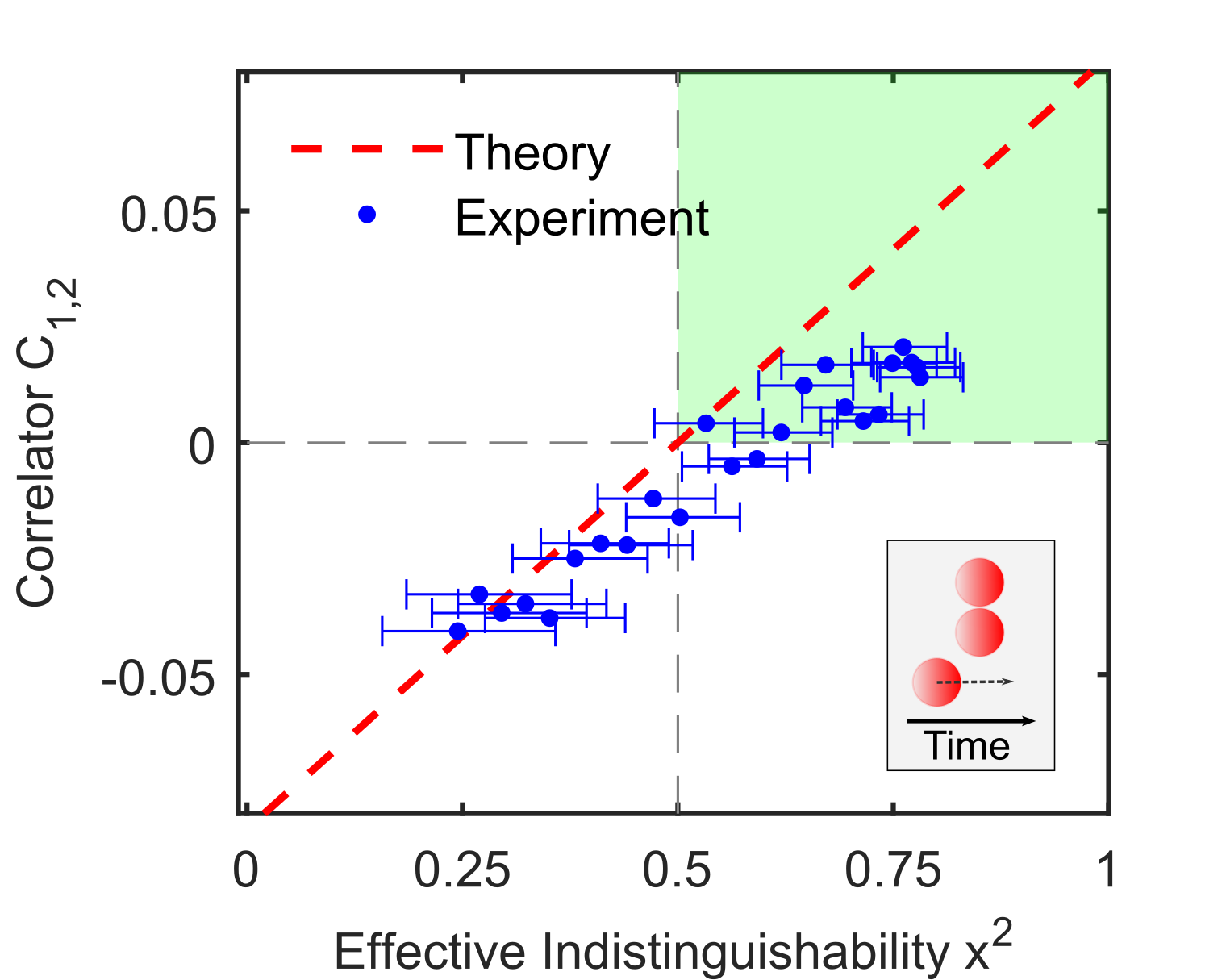}%
    \caption{The correlator for the optimal matrix with imperfect matrix control. The red dashed line indicates the maximal correlator possible with the optimal matrix. The inset shows how the effective indistinguishability was varied by changing temporal overlap of the third photon with respect to the other two photons before injecting them into the chip.}
    \label{fig:figIdeal}
\end{figure}

We begin with examining the statistical distributions of the two-point correlator for Haar-random matrices. We implement $500$ randomly chosen $4 \times 4$ Haar-random matrices. For each implemented matrix, all ${4 \choose 2}=6$  two-mode correlators are measured. Each matrix is measured for 3 minutes \footnote{During data analysis, it became apparent that due to a communication error, that the second phase shifter of the top-left unit cell was not actuated on during measurements. This does not significantly affect the results of the Haar measurements.}.

Figure \ref{fig:figResults}a shows the experimentally observed distributions when two indistinguishable photons are injected into the first two modes of the interferometer. All correlators are negative, as expected, and the distribution follows the expected shape, peaking at the cutoff of $C_c=0$ \cite{Walschaers2016}. The inset shows a zoom-in at the correlators around this cutoff. The overall shape of the distribution is as expected for two photons.

In Fig. \ref{fig:figResults}b, the correlator distribution for 3 fully distinguishable photons is shown. This distribution is similar to that of two photons, but has shifted somewhat towards negative values, As expected, no positive correlator $C > 0$ is observed in this distribution.

Finally, Fig. \ref{fig:figResults}c) shows positive two-point correlators when three partially indistinguishable photons are injected into the network. This is the only case where matrices with $C > 0$ are observed, certifying indistinguishability.

Fig. \ref{fig:figResults}d shows a bar chart of all cases where we observe positive correlators in the experiment with three indistinguishable photons. The label 'ideal' refers to the matrix given in eq. \ref{eq:optimummatrixrows}. The error bars indicate the standard deviation assuming Poissonian noise. Multiple data points, including the one corresponding to $U_{\rm max}$ are statistically significantly above the threshold value $C_c = 0$. The largest observed correlator is $C = 0.027 > C_c$, exceeding the threshold by $7\sigma$. Therefore, we can certify indistinguishability between these three photons.

We now turn out attention exclusively to the matrix $U_\mathrm{max}$. To test the robustness of our witness, we decrease the degree of distinguishability between one of the three photons and the other two by varying the time of arrival in the interferometer. 

Figure \ref{fig:figIdeal} shows the measured value of $C(U_\mathrm{max})$ as a function of $\bar{x}^2$, the average partial distinguishability between the photons. The figure shows that the correlator increases monotonically when the temporal overlap of the photons is increased. The correlator reaches a maximal value $C = 0.0206 \pm\ 0.0038$, as expected. As expected, no data points are observed above the red dashed line when $\bar{x}^2 > 0.5$, which emphasizes the optimal nature of our matrix.

The maximal observed correlator is lower than the expected $C_{q} = 0.052$ for the effective indistinguishability $\bar{x}^2 = 0.81$, and the rate at which $C$ increases with increasing indistinguishability is also lower than expected. Both the reduced slope and correlator can be understood by imperfect chip control. Implementing the optimal matrix with a fidelity of $98$\% already reduces the average correlator to $C = 0.024 \pm 0.007$ for partial indistinguishable photons, where the error bars are over different realizations of the noise. This behaviour is typical for a semi-device-independent witness: imperfect control over the system reduces the parameter regime in which quantumness can be certified, but it never opens up a loophole where quantumness will be claimed if it is not there.

\section{Outlook}
We highlight the fact that it might well be possible to further relax the device independence assumptions introduced here, for example by removing the requirement that the photons are in single-photon Fock states at the input. We also leave the open problem of proving that the conjectured optimal matrix $U_\mathrm{max}$ is in fact optimal.
 
We note that it is straightforward to transfer our scheme into an adversarial setting where one party provides photons which are claimed to be indistinguishable, and the other party either tests this fact or uses those photons to generate a sample from a given boson sampler. 

\section{Conclusion}
In conclusion, we present an experimental demonstration of a semi-device-independent indistinguishability witness for multi-photon interference. The witness requires only the measurement of a single two-point correlator behind two fixed output modes of a programmable optical network. This correlator is believed to be maximized by a given matrix. Consequently, any imperfect implementation of this optimal matrix will inevitably only reduce the correlator, demonstrating its suitability as a semi-device-independent witness. This result opens the way to robust, efficient in-situ characterization of boson sampling experiments. 

\section{Acknowledgements}
\begin{acknowledgments}
We thank Mattia Walschaers for discussions. The Twente team acknowledges funding from the Nederlandse Wetenschaps Organisatie (NWO) via QuantERA QUOMPLEX (Grant No.$~680.91.037$), and Veni (Grant No.$~15872$).
This work has received funding from the European Union’s Horizon 2020 research and innovation programme under grant agreement No 899544.
N.~W.\ acknowledges funding from the European Unions Horizon 2020 research and innovation programme under the Marie Sklodowska-Curie Grant  Agreement No.750905 and the DFG priority
program “Compressed Sensing in Information Processing - Phase 2 (CoSIP2)” and the BMBF 
(QPIC).
\end{acknowledgments}


\begin{thebibliography}{67}%
\makeatletter
\providecommand \@ifxundefined [1]{%
 \@ifx{#1\undefined}
}%
\providecommand \@ifnum [1]{%
 \ifnum #1\expandafter \@firstoftwo
 \else \expandafter \@secondoftwo
 \fi
}%
\providecommand \@ifx [1]{%
 \ifx #1\expandafter \@firstoftwo
 \else \expandafter \@secondoftwo
 \fi
}%
\providecommand \natexlab [1]{#1}%
\providecommand \enquote  [1]{``#1''}%
\providecommand \bibnamefont  [1]{#1}%
\providecommand \bibfnamefont [1]{#1}%
\providecommand \citenamefont [1]{#1}%
\providecommand \href@noop [0]{\@secondoftwo}%
\providecommand \href [0]{\begingroup \@sanitize@url \@href}%
\providecommand \@href[1]{\@@startlink{#1}\@@href}%
\providecommand \@@href[1]{\endgroup#1\@@endlink}%
\providecommand \@sanitize@url [0]{\catcode `\\12\catcode `\$12\catcode
  `\&12\catcode `\#12\catcode `\^12\catcode `\_12\catcode `\%12\relax}%
\providecommand \@@startlink[1]{}%
\providecommand \@@endlink[0]{}%
\providecommand \url  [0]{\begingroup\@sanitize@url \@url }%
\providecommand \@url [1]{\endgroup\@href {#1}{\urlprefix }}%
\providecommand \urlprefix  [0]{URL }%
\providecommand \Eprint [0]{\href }%
\providecommand \doibase [0]{https://doi.org/}%
\providecommand \selectlanguage [0]{\@gobble}%
\providecommand \bibinfo  [0]{\@secondoftwo}%
\providecommand \bibfield  [0]{\@secondoftwo}%
\providecommand \translation [1]{[#1]}%
\providecommand \BibitemOpen [0]{}%
\providecommand \bibitemStop [0]{}%
\providecommand \bibitemNoStop [0]{.\EOS\space}%
\providecommand \EOS [0]{\spacefactor3000\relax}%
\providecommand \BibitemShut  [1]{\csname bibitem#1\endcsname}%
\let\auto@bib@innerbib\@empty
\bibitem [{\citenamefont {Zhong}\ \emph
  {et~al.}(2020{\natexlab{a}})\citenamefont {Zhong}, \citenamefont {Wang},
  \citenamefont {Deng}, \citenamefont {Chen}, \citenamefont {Peng},
  \citenamefont {Luo}, \citenamefont {Qin}, \citenamefont {Wu}, \citenamefont
  {Ding}, \citenamefont {Hu}, \citenamefont {Hu}, \citenamefont {Yang},
  \citenamefont {Zhang}, \citenamefont {Li}, \citenamefont {Li}, \citenamefont
  {Jiang}, \citenamefont {Gan}, \citenamefont {Yang}, \citenamefont {You},
  \citenamefont {Wang}, \citenamefont {Li}, \citenamefont {Liu}, \citenamefont
  {Lu},\ and\ \citenamefont {Pan}}]{zhong_2020_Science}%
  \BibitemOpen
  \bibfield  {author} {\bibinfo {author} {\bibfnamefont {H.-S.}\ \bibnamefont
  {Zhong}}, \bibinfo {author} {\bibfnamefont {H.}~\bibnamefont {Wang}},
  \bibinfo {author} {\bibfnamefont {Y.-H.}\ \bibnamefont {Deng}}, \bibinfo
  {author} {\bibfnamefont {M.-C.}\ \bibnamefont {Chen}}, \bibinfo {author}
  {\bibfnamefont {L.-C.}\ \bibnamefont {Peng}}, \bibinfo {author}
  {\bibfnamefont {Y.-H.}\ \bibnamefont {Luo}}, \bibinfo {author} {\bibfnamefont
  {J.}~\bibnamefont {Qin}}, \bibinfo {author} {\bibfnamefont {D.}~\bibnamefont
  {Wu}}, \bibinfo {author} {\bibfnamefont {X.}~\bibnamefont {Ding}}, \bibinfo
  {author} {\bibfnamefont {Y.}~\bibnamefont {Hu}}, \bibinfo {author}
  {\bibfnamefont {P.}~\bibnamefont {Hu}}, \bibinfo {author} {\bibfnamefont
  {X.-Y.}\ \bibnamefont {Yang}}, \bibinfo {author} {\bibfnamefont {W.-J.}\
  \bibnamefont {Zhang}}, \bibinfo {author} {\bibfnamefont {H.}~\bibnamefont
  {Li}}, \bibinfo {author} {\bibfnamefont {Y.}~\bibnamefont {Li}}, \bibinfo
  {author} {\bibfnamefont {X.}~\bibnamefont {Jiang}}, \bibinfo {author}
  {\bibfnamefont {L.}~\bibnamefont {Gan}}, \bibinfo {author} {\bibfnamefont
  {G.}~\bibnamefont {Yang}}, \bibinfo {author} {\bibfnamefont {L.}~\bibnamefont
  {You}}, \bibinfo {author} {\bibfnamefont {Z.}~\bibnamefont {Wang}}, \bibinfo
  {author} {\bibfnamefont {L.}~\bibnamefont {Li}}, \bibinfo {author}
  {\bibfnamefont {N.-L.}\ \bibnamefont {Liu}}, \bibinfo {author} {\bibfnamefont
  {C.-Y.}\ \bibnamefont {Lu}},\ and\ \bibinfo {author} {\bibfnamefont {J.-W.}\
  \bibnamefont {Pan}},\ }\bibfield  {journal} {\bibinfo  {journal} {Science}\
  }\href {https://doi.org/10.1126/science.abe8770} {10.1126/science.abe8770}
  (\bibinfo {year} {2020}{\natexlab{a}})\BibitemShut {NoStop}%
\bibitem [{\citenamefont {Zhong}\ \emph
  {et~al.}(2021{\natexlab{a}})\citenamefont {Zhong}, \citenamefont {Deng},
  \citenamefont {Qin}, \citenamefont {Wang}, \citenamefont {Chen},
  \citenamefont {Peng}, \citenamefont {Luo}, \citenamefont {Wu}, \citenamefont
  {Gong}, \citenamefont {Su}, \citenamefont {Hu}, \citenamefont {Hu},
  \citenamefont {Yang}, \citenamefont {Zhang}, \citenamefont {Li},
  \citenamefont {Li}, \citenamefont {Jiang}, \citenamefont {Gan}, \citenamefont
  {Yang}, \citenamefont {You}, \citenamefont {Wang}, \citenamefont {Li},
  \citenamefont {Liu}, \citenamefont {Renema}, \citenamefont {Lu},\ and\
  \citenamefont {Pan}}]{zhong_2021_ArXiv}%
  \BibitemOpen
  \bibfield  {author} {\bibinfo {author} {\bibfnamefont {H.-S.}\ \bibnamefont
  {Zhong}}, \bibinfo {author} {\bibfnamefont {Y.-H.}\ \bibnamefont {Deng}},
  \bibinfo {author} {\bibfnamefont {J.}~\bibnamefont {Qin}}, \bibinfo {author}
  {\bibfnamefont {H.}~\bibnamefont {Wang}}, \bibinfo {author} {\bibfnamefont
  {M.-C.}\ \bibnamefont {Chen}}, \bibinfo {author} {\bibfnamefont {L.-C.}\
  \bibnamefont {Peng}}, \bibinfo {author} {\bibfnamefont {Y.-H.}\ \bibnamefont
  {Luo}}, \bibinfo {author} {\bibfnamefont {D.}~\bibnamefont {Wu}}, \bibinfo
  {author} {\bibfnamefont {S.-Q.}\ \bibnamefont {Gong}}, \bibinfo {author}
  {\bibfnamefont {H.}~\bibnamefont {Su}}, \bibinfo {author} {\bibfnamefont
  {Y.}~\bibnamefont {Hu}}, \bibinfo {author} {\bibfnamefont {P.}~\bibnamefont
  {Hu}}, \bibinfo {author} {\bibfnamefont {X.-Y.}\ \bibnamefont {Yang}},
  \bibinfo {author} {\bibfnamefont {W.-J.}\ \bibnamefont {Zhang}}, \bibinfo
  {author} {\bibfnamefont {H.}~\bibnamefont {Li}}, \bibinfo {author}
  {\bibfnamefont {Y.}~\bibnamefont {Li}}, \bibinfo {author} {\bibfnamefont
  {X.}~\bibnamefont {Jiang}}, \bibinfo {author} {\bibfnamefont
  {L.}~\bibnamefont {Gan}}, \bibinfo {author} {\bibfnamefont {G.}~\bibnamefont
  {Yang}}, \bibinfo {author} {\bibfnamefont {L.}~\bibnamefont {You}}, \bibinfo
  {author} {\bibfnamefont {Z.}~\bibnamefont {Wang}}, \bibinfo {author}
  {\bibfnamefont {L.}~\bibnamefont {Li}}, \bibinfo {author} {\bibfnamefont
  {N.-L.}\ \bibnamefont {Liu}}, \bibinfo {author} {\bibfnamefont
  {J.}~\bibnamefont {Renema}}, \bibinfo {author} {\bibfnamefont {C.-Y.}\
  \bibnamefont {Lu}},\ and\ \bibinfo {author} {\bibfnamefont {J.-W.}\
  \bibnamefont {Pan}},\ }\href@noop {} {\bibfield  {journal} {\bibinfo
  {journal} {arXiv:2106.15534 [physics, physics:quant-ph]}\ } (\bibinfo {year}
  {2021}{\natexlab{a}})},\ \Eprint {https://arxiv.org/abs/2106.15534}
  {arXiv:2106.15534 [physics, physics:quant-ph]} \BibitemShut {NoStop}%
\bibitem [{\citenamefont {Broome}\ \emph {et~al.}(2013)\citenamefont {Broome},
  \citenamefont {Fedrizzi}, \citenamefont {{Rahimi-Keshari}}, \citenamefont
  {Dove}, \citenamefont {Aaronson}, \citenamefont {Ralph},\ and\ \citenamefont
  {White}}]{broome_2013_Science}%
  \BibitemOpen
  \bibfield  {author} {\bibinfo {author} {\bibfnamefont {M.~A.}\ \bibnamefont
  {Broome}}, \bibinfo {author} {\bibfnamefont {A.}~\bibnamefont {Fedrizzi}},
  \bibinfo {author} {\bibfnamefont {S.}~\bibnamefont {{Rahimi-Keshari}}},
  \bibinfo {author} {\bibfnamefont {J.}~\bibnamefont {Dove}}, \bibinfo {author}
  {\bibfnamefont {S.}~\bibnamefont {Aaronson}}, \bibinfo {author}
  {\bibfnamefont {T.~C.}\ \bibnamefont {Ralph}},\ and\ \bibinfo {author}
  {\bibfnamefont {A.~G.}\ \bibnamefont {White}},\ }\href
  {https://doi.org/10.1126/science.1231440} {\bibfield  {journal} {\bibinfo
  {journal} {Science}\ }\textbf {\bibinfo {volume} {339}},\ \bibinfo {pages}
  {794} (\bibinfo {year} {2013})}\BibitemShut {NoStop}%
\bibitem [{\citenamefont {Spring}\ \emph {et~al.}(2013)\citenamefont {Spring},
  \citenamefont {Metcalf}, \citenamefont {Humphreys}, \citenamefont
  {Kolthammer}, \citenamefont {Jin}, \citenamefont {Barbieri}, \citenamefont
  {Datta}, \citenamefont {{Thomas-Peter}}, \citenamefont {Langford},
  \citenamefont {Kundys}, \citenamefont {Gates}, \citenamefont {Smith},
  \citenamefont {Smith},\ and\ \citenamefont {Walmsley}}]{spring_2013_Science}%
  \BibitemOpen
  \bibfield  {author} {\bibinfo {author} {\bibfnamefont {J.~B.}\ \bibnamefont
  {Spring}}, \bibinfo {author} {\bibfnamefont {B.~J.}\ \bibnamefont {Metcalf}},
  \bibinfo {author} {\bibfnamefont {P.~C.}\ \bibnamefont {Humphreys}}, \bibinfo
  {author} {\bibfnamefont {W.~S.}\ \bibnamefont {Kolthammer}}, \bibinfo
  {author} {\bibfnamefont {X.-M.}\ \bibnamefont {Jin}}, \bibinfo {author}
  {\bibfnamefont {M.}~\bibnamefont {Barbieri}}, \bibinfo {author}
  {\bibfnamefont {A.}~\bibnamefont {Datta}}, \bibinfo {author} {\bibfnamefont
  {N.}~\bibnamefont {{Thomas-Peter}}}, \bibinfo {author} {\bibfnamefont
  {N.~K.}\ \bibnamefont {Langford}}, \bibinfo {author} {\bibfnamefont
  {D.}~\bibnamefont {Kundys}}, \bibinfo {author} {\bibfnamefont {J.~C.}\
  \bibnamefont {Gates}}, \bibinfo {author} {\bibfnamefont {B.~J.}\ \bibnamefont
  {Smith}}, \bibinfo {author} {\bibfnamefont {P.~G.~R.}\ \bibnamefont
  {Smith}},\ and\ \bibinfo {author} {\bibfnamefont {I.~A.}\ \bibnamefont
  {Walmsley}},\ }\href {https://doi.org/10.1126/science.1231692} {\bibfield
  {journal} {\bibinfo  {journal} {Science}\ }\textbf {\bibinfo {volume}
  {339}},\ \bibinfo {pages} {798} (\bibinfo {year} {2013})}\BibitemShut
  {NoStop}%
\bibitem [{\citenamefont {Tillmann}\ \emph {et~al.}(2013)\citenamefont
  {Tillmann}, \citenamefont {Daki{\'c}}, \citenamefont {Heilmann},
  \citenamefont {Nolte}, \citenamefont {Szameit},\ and\ \citenamefont
  {Walther}}]{tillmann_2013_NatPhotonics}%
  \BibitemOpen
  \bibfield  {author} {\bibinfo {author} {\bibfnamefont {M.}~\bibnamefont
  {Tillmann}}, \bibinfo {author} {\bibfnamefont {B.}~\bibnamefont {Daki{\'c}}},
  \bibinfo {author} {\bibfnamefont {R.}~\bibnamefont {Heilmann}}, \bibinfo
  {author} {\bibfnamefont {S.}~\bibnamefont {Nolte}}, \bibinfo {author}
  {\bibfnamefont {A.}~\bibnamefont {Szameit}},\ and\ \bibinfo {author}
  {\bibfnamefont {P.}~\bibnamefont {Walther}},\ }\href
  {https://doi.org/10.1038/nphoton.2013.102} {\bibfield  {journal} {\bibinfo
  {journal} {Nat. Photonics}\ }\textbf {\bibinfo {volume} {7}},\ \bibinfo
  {pages} {540} (\bibinfo {year} {2013})}\BibitemShut {NoStop}%
\bibitem [{\citenamefont {Crespi}\ \emph {et~al.}(2013)\citenamefont {Crespi},
  \citenamefont {Osellame}, \citenamefont {Ramponi}, \citenamefont {Brod},
  \citenamefont {Galv{\~a}o}, \citenamefont {Spagnolo}, \citenamefont
  {Vitelli}, \citenamefont {Maiorino}, \citenamefont {Mataloni},\ and\
  \citenamefont {Sciarrino}}]{crespi_2013_NatPhotonics}%
  \BibitemOpen
  \bibfield  {author} {\bibinfo {author} {\bibfnamefont {A.}~\bibnamefont
  {Crespi}}, \bibinfo {author} {\bibfnamefont {R.}~\bibnamefont {Osellame}},
  \bibinfo {author} {\bibfnamefont {R.}~\bibnamefont {Ramponi}}, \bibinfo
  {author} {\bibfnamefont {D.~J.}\ \bibnamefont {Brod}}, \bibinfo {author}
  {\bibfnamefont {E.~F.}\ \bibnamefont {Galv{\~a}o}}, \bibinfo {author}
  {\bibfnamefont {N.}~\bibnamefont {Spagnolo}}, \bibinfo {author}
  {\bibfnamefont {C.}~\bibnamefont {Vitelli}}, \bibinfo {author} {\bibfnamefont
  {E.}~\bibnamefont {Maiorino}}, \bibinfo {author} {\bibfnamefont
  {P.}~\bibnamefont {Mataloni}},\ and\ \bibinfo {author} {\bibfnamefont
  {F.}~\bibnamefont {Sciarrino}},\ }\href
  {https://doi.org/10.1038/nphoton.2013.112} {\bibfield  {journal} {\bibinfo
  {journal} {Nat. Photonics}\ }\textbf {\bibinfo {volume} {7}},\ \bibinfo
  {pages} {545} (\bibinfo {year} {2013})}\BibitemShut {NoStop}%
\bibitem [{\citenamefont {Wang}\ \emph {et~al.}(2017)\citenamefont {Wang},
  \citenamefont {He}, \citenamefont {Li}, \citenamefont {Su}, \citenamefont
  {Li}, \citenamefont {Huang}, \citenamefont {Ding}, \citenamefont {Chen},
  \citenamefont {Liu}, \citenamefont {Qin}, \citenamefont {Li}, \citenamefont
  {He}, \citenamefont {Schneider}, \citenamefont {Kamp}, \citenamefont {Peng},
  \citenamefont {H{\"o}fling}, \citenamefont {Lu},\ and\ \citenamefont
  {Pan}}]{wang_2017_Nat.Photonics}%
  \BibitemOpen
  \bibfield  {author} {\bibinfo {author} {\bibfnamefont {H.}~\bibnamefont
  {Wang}}, \bibinfo {author} {\bibfnamefont {Y.}~\bibnamefont {He}}, \bibinfo
  {author} {\bibfnamefont {Y.-H.}\ \bibnamefont {Li}}, \bibinfo {author}
  {\bibfnamefont {Z.-E.}\ \bibnamefont {Su}}, \bibinfo {author} {\bibfnamefont
  {B.}~\bibnamefont {Li}}, \bibinfo {author} {\bibfnamefont {H.-L.}\
  \bibnamefont {Huang}}, \bibinfo {author} {\bibfnamefont {X.}~\bibnamefont
  {Ding}}, \bibinfo {author} {\bibfnamefont {M.-C.}\ \bibnamefont {Chen}},
  \bibinfo {author} {\bibfnamefont {C.}~\bibnamefont {Liu}}, \bibinfo {author}
  {\bibfnamefont {J.}~\bibnamefont {Qin}}, \bibinfo {author} {\bibfnamefont
  {J.-P.}\ \bibnamefont {Li}}, \bibinfo {author} {\bibfnamefont {Y.-M.}\
  \bibnamefont {He}}, \bibinfo {author} {\bibfnamefont {C.}~\bibnamefont
  {Schneider}}, \bibinfo {author} {\bibfnamefont {M.}~\bibnamefont {Kamp}},
  \bibinfo {author} {\bibfnamefont {C.-Z.}\ \bibnamefont {Peng}}, \bibinfo
  {author} {\bibfnamefont {S.}~\bibnamefont {H{\"o}fling}}, \bibinfo {author}
  {\bibfnamefont {C.-Y.}\ \bibnamefont {Lu}},\ and\ \bibinfo {author}
  {\bibfnamefont {J.-W.}\ \bibnamefont {Pan}},\ }\href
  {https://doi.org/10.1038/nphoton.2017.63} {\bibfield  {journal} {\bibinfo
  {journal} {Nat. Photonics}\ }\textbf {\bibinfo {volume} {11}},\ \bibinfo
  {pages} {361} (\bibinfo {year} {2017})}\BibitemShut {NoStop}%
\bibitem [{\citenamefont {He}\ \emph {et~al.}(2017)\citenamefont {He},
  \citenamefont {Ding}, \citenamefont {Su}, \citenamefont {Huang},
  \citenamefont {Qin}, \citenamefont {Wang}, \citenamefont {Unsleber},
  \citenamefont {Chen}, \citenamefont {Wang}, \citenamefont {He}, \citenamefont
  {Wang}, \citenamefont {Zhang}, \citenamefont {Chen}, \citenamefont
  {Schneider}, \citenamefont {Kamp}, \citenamefont {You}, \citenamefont {Wang},
  \citenamefont {H{\"o}fling}, \citenamefont {Lu},\ and\ \citenamefont
  {Pan}}]{He2017}%
  \BibitemOpen
  \bibfield  {author} {\bibinfo {author} {\bibfnamefont {Y.}~\bibnamefont
  {He}}, \bibinfo {author} {\bibfnamefont {X.}~\bibnamefont {Ding}}, \bibinfo
  {author} {\bibfnamefont {Z.-E.}\ \bibnamefont {Su}}, \bibinfo {author}
  {\bibfnamefont {H.-L.}\ \bibnamefont {Huang}}, \bibinfo {author}
  {\bibfnamefont {J.}~\bibnamefont {Qin}}, \bibinfo {author} {\bibfnamefont
  {C.}~\bibnamefont {Wang}}, \bibinfo {author} {\bibfnamefont {S.}~\bibnamefont
  {Unsleber}}, \bibinfo {author} {\bibfnamefont {C.}~\bibnamefont {Chen}},
  \bibinfo {author} {\bibfnamefont {H.}~\bibnamefont {Wang}}, \bibinfo {author}
  {\bibfnamefont {Y.-M.}\ \bibnamefont {He}}, \bibinfo {author} {\bibfnamefont
  {X.-L.}\ \bibnamefont {Wang}}, \bibinfo {author} {\bibfnamefont {W.-J.}\
  \bibnamefont {Zhang}}, \bibinfo {author} {\bibfnamefont {S.-J.}\ \bibnamefont
  {Chen}}, \bibinfo {author} {\bibfnamefont {C.}~\bibnamefont {Schneider}},
  \bibinfo {author} {\bibfnamefont {M.}~\bibnamefont {Kamp}}, \bibinfo {author}
  {\bibfnamefont {L.-X.}\ \bibnamefont {You}}, \bibinfo {author} {\bibfnamefont
  {Z.}~\bibnamefont {Wang}}, \bibinfo {author} {\bibfnamefont {S.}~\bibnamefont
  {H{\"o}fling}}, \bibinfo {author} {\bibfnamefont {C.-Y.}\ \bibnamefont
  {Lu}},\ and\ \bibinfo {author} {\bibfnamefont {J.-W.}\ \bibnamefont {Pan}},\
  }\href {https://doi.org/10.1103/PhysRevLett.118.190501} {\bibfield  {journal}
  {\bibinfo  {journal} {Phys. Rev. Lett.}\ }\textbf {\bibinfo {volume} {118}},\
  \bibinfo {pages} {190501} (\bibinfo {year} {2017})}\BibitemShut {NoStop}%
\bibitem [{\citenamefont {Paesani}\ \emph {et~al.}(2019)\citenamefont
  {Paesani}, \citenamefont {Ding}, \citenamefont {Santagati}, \citenamefont
  {Chakhmakhchyan}, \citenamefont {Vigliar}, \citenamefont {Rottwitt},
  \citenamefont {Oxenl{\o}we}, \citenamefont {Wang}, \citenamefont {Thompson},\
  and\ \citenamefont {Laing}}]{Paesani2019}%
  \BibitemOpen
  \bibfield  {author} {\bibinfo {author} {\bibfnamefont {S.}~\bibnamefont
  {Paesani}}, \bibinfo {author} {\bibfnamefont {Y.}~\bibnamefont {Ding}},
  \bibinfo {author} {\bibfnamefont {R.}~\bibnamefont {Santagati}}, \bibinfo
  {author} {\bibfnamefont {L.}~\bibnamefont {Chakhmakhchyan}}, \bibinfo
  {author} {\bibfnamefont {C.}~\bibnamefont {Vigliar}}, \bibinfo {author}
  {\bibfnamefont {K.}~\bibnamefont {Rottwitt}}, \bibinfo {author}
  {\bibfnamefont {L.~K.}\ \bibnamefont {Oxenl{\o}we}}, \bibinfo {author}
  {\bibfnamefont {J.}~\bibnamefont {Wang}}, \bibinfo {author} {\bibfnamefont
  {M.~G.}\ \bibnamefont {Thompson}},\ and\ \bibinfo {author} {\bibfnamefont
  {A.}~\bibnamefont {Laing}},\ }\href
  {https://doi.org/10.1038/s41567-019-0567-8} {\bibfield  {journal} {\bibinfo
  {journal} {Nat. Phys.}\ }\textbf {\bibinfo {volume} {15}},\ \bibinfo {pages}
  {925} (\bibinfo {year} {2019})}\BibitemShut {NoStop}%
\bibitem [{\citenamefont {Zhong}\ \emph
  {et~al.}(2020{\natexlab{b}})\citenamefont {Zhong}, \citenamefont {Wang},
  \citenamefont {Deng}, \citenamefont {Chen}, \citenamefont {Peng},
  \citenamefont {Luo}, \citenamefont {Qin}, \citenamefont {Wu}, \citenamefont
  {Ding}, \citenamefont {Hu}, \citenamefont {Hu}, \citenamefont {Yang},
  \citenamefont {Zhang}, \citenamefont {Li}, \citenamefont {Li}, \citenamefont
  {Jiang}, \citenamefont {Gan}, \citenamefont {Yang}, \citenamefont {You},
  \citenamefont {Wang}, \citenamefont {Li}, \citenamefont {Liu}, \citenamefont
  {Lu},\ and\ \citenamefont {Pan}}]{Zhong2020}%
  \BibitemOpen
  \bibfield  {author} {\bibinfo {author} {\bibfnamefont {H.-S.}\ \bibnamefont
  {Zhong}}, \bibinfo {author} {\bibfnamefont {H.}~\bibnamefont {Wang}},
  \bibinfo {author} {\bibfnamefont {Y.-H.}\ \bibnamefont {Deng}}, \bibinfo
  {author} {\bibfnamefont {M.-C.}\ \bibnamefont {Chen}}, \bibinfo {author}
  {\bibfnamefont {L.-C.}\ \bibnamefont {Peng}}, \bibinfo {author}
  {\bibfnamefont {Y.-H.}\ \bibnamefont {Luo}}, \bibinfo {author} {\bibfnamefont
  {J.}~\bibnamefont {Qin}}, \bibinfo {author} {\bibfnamefont {D.}~\bibnamefont
  {Wu}}, \bibinfo {author} {\bibfnamefont {X.}~\bibnamefont {Ding}}, \bibinfo
  {author} {\bibfnamefont {Y.}~\bibnamefont {Hu}}, \bibinfo {author}
  {\bibfnamefont {P.}~\bibnamefont {Hu}}, \bibinfo {author} {\bibfnamefont
  {X.-Y.}\ \bibnamefont {Yang}}, \bibinfo {author} {\bibfnamefont {W.-J.}\
  \bibnamefont {Zhang}}, \bibinfo {author} {\bibfnamefont {H.}~\bibnamefont
  {Li}}, \bibinfo {author} {\bibfnamefont {Y.}~\bibnamefont {Li}}, \bibinfo
  {author} {\bibfnamefont {X.}~\bibnamefont {Jiang}}, \bibinfo {author}
  {\bibfnamefont {L.}~\bibnamefont {Gan}}, \bibinfo {author} {\bibfnamefont
  {G.}~\bibnamefont {Yang}}, \bibinfo {author} {\bibfnamefont {L.}~\bibnamefont
  {You}}, \bibinfo {author} {\bibfnamefont {Z.}~\bibnamefont {Wang}}, \bibinfo
  {author} {\bibfnamefont {L.}~\bibnamefont {Li}}, \bibinfo {author}
  {\bibfnamefont {N.-L.}\ \bibnamefont {Liu}}, \bibinfo {author} {\bibfnamefont
  {C.-Y.}\ \bibnamefont {Lu}},\ and\ \bibinfo {author} {\bibfnamefont {J.-W.}\
  \bibnamefont {Pan}},\ }\bibfield  {journal} {\bibinfo  {journal} {Science}\
  }\href {https://doi.org/10.1126/science.abe8770} {10.1126/science.abe8770}
  (\bibinfo {year} {2020}{\natexlab{b}})\BibitemShut {NoStop}%
\bibitem [{\citenamefont {Zhong}\ \emph
  {et~al.}(2021{\natexlab{b}})\citenamefont {Zhong}, \citenamefont {Deng},
  \citenamefont {Qin}, \citenamefont {Wang}, \citenamefont {Chen},
  \citenamefont {Peng}, \citenamefont {Luo}, \citenamefont {Wu}, \citenamefont
  {Gong}, \citenamefont {Su}, \citenamefont {Hu}, \citenamefont {Hu},
  \citenamefont {Yang}, \citenamefont {Zhang}, \citenamefont {Li},
  \citenamefont {Li}, \citenamefont {Jiang}, \citenamefont {Gan}, \citenamefont
  {Yang}, \citenamefont {You}, \citenamefont {Wang}, \citenamefont {Li},
  \citenamefont {Liu}, \citenamefont {Renema}, \citenamefont {Lu},\ and\
  \citenamefont {Pan}}]{Zhong2021}%
  \BibitemOpen
  \bibfield  {author} {\bibinfo {author} {\bibfnamefont {H.-S.}\ \bibnamefont
  {Zhong}}, \bibinfo {author} {\bibfnamefont {Y.-H.}\ \bibnamefont {Deng}},
  \bibinfo {author} {\bibfnamefont {J.}~\bibnamefont {Qin}}, \bibinfo {author}
  {\bibfnamefont {H.}~\bibnamefont {Wang}}, \bibinfo {author} {\bibfnamefont
  {M.-C.}\ \bibnamefont {Chen}}, \bibinfo {author} {\bibfnamefont {L.-C.}\
  \bibnamefont {Peng}}, \bibinfo {author} {\bibfnamefont {Y.-H.}\ \bibnamefont
  {Luo}}, \bibinfo {author} {\bibfnamefont {D.}~\bibnamefont {Wu}}, \bibinfo
  {author} {\bibfnamefont {S.-Q.}\ \bibnamefont {Gong}}, \bibinfo {author}
  {\bibfnamefont {H.}~\bibnamefont {Su}}, \bibinfo {author} {\bibfnamefont
  {Y.}~\bibnamefont {Hu}}, \bibinfo {author} {\bibfnamefont {P.}~\bibnamefont
  {Hu}}, \bibinfo {author} {\bibfnamefont {X.-Y.}\ \bibnamefont {Yang}},
  \bibinfo {author} {\bibfnamefont {W.-J.}\ \bibnamefont {Zhang}}, \bibinfo
  {author} {\bibfnamefont {H.}~\bibnamefont {Li}}, \bibinfo {author}
  {\bibfnamefont {Y.}~\bibnamefont {Li}}, \bibinfo {author} {\bibfnamefont
  {X.}~\bibnamefont {Jiang}}, \bibinfo {author} {\bibfnamefont
  {L.}~\bibnamefont {Gan}}, \bibinfo {author} {\bibfnamefont {G.}~\bibnamefont
  {Yang}}, \bibinfo {author} {\bibfnamefont {L.}~\bibnamefont {You}}, \bibinfo
  {author} {\bibfnamefont {Z.}~\bibnamefont {Wang}}, \bibinfo {author}
  {\bibfnamefont {L.}~\bibnamefont {Li}}, \bibinfo {author} {\bibfnamefont
  {N.-L.}\ \bibnamefont {Liu}}, \bibinfo {author} {\bibfnamefont
  {J.}~\bibnamefont {Renema}}, \bibinfo {author} {\bibfnamefont {C.-Y.}\
  \bibnamefont {Lu}},\ and\ \bibinfo {author} {\bibfnamefont {J.-W.}\
  \bibnamefont {Pan}},\ }\href@noop {} {\bibfield  {journal} {\bibinfo
  {journal} {arXiv:2106.15534 [physics, physics:quant-ph]}\ } (\bibinfo {year}
  {2021}{\natexlab{b}})},\ \Eprint {https://arxiv.org/abs/2106.15534}
  {arXiv:2106.15534 [physics, physics:quant-ph]} \BibitemShut {NoStop}%
\bibitem [{\citenamefont {Arrazola}\ \emph {et~al.}(2021)\citenamefont
  {Arrazola}, \citenamefont {Bergholm}, \citenamefont {Br{\'a}dler},
  \citenamefont {Bromley}, \citenamefont {Collins}, \citenamefont {Dhand},
  \citenamefont {Fumagalli}, \citenamefont {Gerrits}, \citenamefont {Goussev},
  \citenamefont {Helt}, \citenamefont {Hundal}, \citenamefont {Isacsson},
  \citenamefont {Israel}, \citenamefont {Izaac}, \citenamefont {Jahangiri},
  \citenamefont {Janik}, \citenamefont {Killoran}, \citenamefont {Kumar},
  \citenamefont {Lavoie}, \citenamefont {Lita}, \citenamefont {Mahler},
  \citenamefont {Menotti}, \citenamefont {Morrison}, \citenamefont {Nam},
  \citenamefont {Neuhaus}, \citenamefont {Qi}, \citenamefont {Quesada},
  \citenamefont {Repingon}, \citenamefont {Sabapathy}, \citenamefont {Schuld},
  \citenamefont {Su}, \citenamefont {Swinarton}, \citenamefont {Sz{\'a}va},
  \citenamefont {Tan}, \citenamefont {Tan}, \citenamefont {Vaidya},
  \citenamefont {Vernon}, \citenamefont {Zabaneh},\ and\ \citenamefont
  {Zhang}}]{Arrazola2021}%
  \BibitemOpen
  \bibfield  {author} {\bibinfo {author} {\bibfnamefont {J.~M.}\ \bibnamefont
  {Arrazola}}, \bibinfo {author} {\bibfnamefont {V.}~\bibnamefont {Bergholm}},
  \bibinfo {author} {\bibfnamefont {K.}~\bibnamefont {Br{\'a}dler}}, \bibinfo
  {author} {\bibfnamefont {T.~R.}\ \bibnamefont {Bromley}}, \bibinfo {author}
  {\bibfnamefont {M.~J.}\ \bibnamefont {Collins}}, \bibinfo {author}
  {\bibfnamefont {I.}~\bibnamefont {Dhand}}, \bibinfo {author} {\bibfnamefont
  {A.}~\bibnamefont {Fumagalli}}, \bibinfo {author} {\bibfnamefont
  {T.}~\bibnamefont {Gerrits}}, \bibinfo {author} {\bibfnamefont
  {A.}~\bibnamefont {Goussev}}, \bibinfo {author} {\bibfnamefont {L.~G.}\
  \bibnamefont {Helt}}, \bibinfo {author} {\bibfnamefont {J.}~\bibnamefont
  {Hundal}}, \bibinfo {author} {\bibfnamefont {T.}~\bibnamefont {Isacsson}},
  \bibinfo {author} {\bibfnamefont {R.~B.}\ \bibnamefont {Israel}}, \bibinfo
  {author} {\bibfnamefont {J.}~\bibnamefont {Izaac}}, \bibinfo {author}
  {\bibfnamefont {S.}~\bibnamefont {Jahangiri}}, \bibinfo {author}
  {\bibfnamefont {R.}~\bibnamefont {Janik}}, \bibinfo {author} {\bibfnamefont
  {N.}~\bibnamefont {Killoran}}, \bibinfo {author} {\bibfnamefont {S.~P.}\
  \bibnamefont {Kumar}}, \bibinfo {author} {\bibfnamefont {J.}~\bibnamefont
  {Lavoie}}, \bibinfo {author} {\bibfnamefont {A.~E.}\ \bibnamefont {Lita}},
  \bibinfo {author} {\bibfnamefont {D.~H.}\ \bibnamefont {Mahler}}, \bibinfo
  {author} {\bibfnamefont {M.}~\bibnamefont {Menotti}}, \bibinfo {author}
  {\bibfnamefont {B.}~\bibnamefont {Morrison}}, \bibinfo {author}
  {\bibfnamefont {S.~W.}\ \bibnamefont {Nam}}, \bibinfo {author} {\bibfnamefont
  {L.}~\bibnamefont {Neuhaus}}, \bibinfo {author} {\bibfnamefont {H.~Y.}\
  \bibnamefont {Qi}}, \bibinfo {author} {\bibfnamefont {N.}~\bibnamefont
  {Quesada}}, \bibinfo {author} {\bibfnamefont {A.}~\bibnamefont {Repingon}},
  \bibinfo {author} {\bibfnamefont {K.~K.}\ \bibnamefont {Sabapathy}}, \bibinfo
  {author} {\bibfnamefont {M.}~\bibnamefont {Schuld}}, \bibinfo {author}
  {\bibfnamefont {D.}~\bibnamefont {Su}}, \bibinfo {author} {\bibfnamefont
  {J.}~\bibnamefont {Swinarton}}, \bibinfo {author} {\bibfnamefont
  {A.}~\bibnamefont {Sz{\'a}va}}, \bibinfo {author} {\bibfnamefont
  {K.}~\bibnamefont {Tan}}, \bibinfo {author} {\bibfnamefont {P.}~\bibnamefont
  {Tan}}, \bibinfo {author} {\bibfnamefont {V.~D.}\ \bibnamefont {Vaidya}},
  \bibinfo {author} {\bibfnamefont {Z.}~\bibnamefont {Vernon}}, \bibinfo
  {author} {\bibfnamefont {Z.}~\bibnamefont {Zabaneh}},\ and\ \bibinfo {author}
  {\bibfnamefont {Y.}~\bibnamefont {Zhang}},\ }\href
  {https://doi.org/10.1038/s41586-021-03202-1} {\bibfield  {journal} {\bibinfo
  {journal} {Nature}\ }\textbf {\bibinfo {volume} {591}},\ \bibinfo {pages}
  {54} (\bibinfo {year} {2021})}\BibitemShut {NoStop}%
\bibitem [{\citenamefont {Aaronson}\ and\ \citenamefont
  {Arkhipov}(2013)}]{aaronson_2013_TheoryComputa}%
  \BibitemOpen
  \bibfield  {author} {\bibinfo {author} {\bibfnamefont {S.}~\bibnamefont
  {Aaronson}}\ and\ \bibinfo {author} {\bibfnamefont {A.}~\bibnamefont
  {Arkhipov}},\ }\href@noop {} {\bibfield  {journal} {\bibinfo  {journal}
  {Theory Comput.}\ }\textbf {\bibinfo {volume} {9}},\ \bibinfo {pages} {143}
  (\bibinfo {year} {2013})}\BibitemShut {NoStop}%
\bibitem [{\citenamefont {Deshpande}\ \emph {et~al.}(2021)\citenamefont
  {Deshpande}, \citenamefont {Mehta}, \citenamefont {Vincent}, \citenamefont
  {Quesada}, \citenamefont {Hinsche}, \citenamefont {Ioannou}, \citenamefont
  {Madsen}, \citenamefont {Lavoie}, \citenamefont {Qi}, \citenamefont {Eisert},
  \citenamefont {Hangleiter}, \citenamefont {Fefferman},\ and\ \citenamefont
  {Dhand}}]{deshpande_2021_ArXiv}%
  \BibitemOpen
  \bibfield  {author} {\bibinfo {author} {\bibfnamefont {A.}~\bibnamefont
  {Deshpande}}, \bibinfo {author} {\bibfnamefont {A.}~\bibnamefont {Mehta}},
  \bibinfo {author} {\bibfnamefont {T.}~\bibnamefont {Vincent}}, \bibinfo
  {author} {\bibfnamefont {N.}~\bibnamefont {Quesada}}, \bibinfo {author}
  {\bibfnamefont {M.}~\bibnamefont {Hinsche}}, \bibinfo {author} {\bibfnamefont
  {M.}~\bibnamefont {Ioannou}}, \bibinfo {author} {\bibfnamefont
  {L.}~\bibnamefont {Madsen}}, \bibinfo {author} {\bibfnamefont
  {J.}~\bibnamefont {Lavoie}}, \bibinfo {author} {\bibfnamefont
  {H.}~\bibnamefont {Qi}}, \bibinfo {author} {\bibfnamefont {J.}~\bibnamefont
  {Eisert}}, \bibinfo {author} {\bibfnamefont {D.}~\bibnamefont {Hangleiter}},
  \bibinfo {author} {\bibfnamefont {B.}~\bibnamefont {Fefferman}},\ and\
  \bibinfo {author} {\bibfnamefont {I.}~\bibnamefont {Dhand}},\ }\href@noop {}
  {\bibfield  {journal} {\bibinfo  {journal} {arXiv:2102.12474 [quant-ph]}\ }
  (\bibinfo {year} {2021})},\ \Eprint {https://arxiv.org/abs/2102.12474}
  {arXiv:2102.12474 [quant-ph]} \BibitemShut {NoStop}%
\bibitem [{\citenamefont {Renema}\ \emph
  {et~al.}(2018{\natexlab{a}})\citenamefont {Renema}, \citenamefont {Menssen},
  \citenamefont {Clements}, \citenamefont {Triginer}, \citenamefont
  {Kolthammer},\ and\ \citenamefont {Walmsley}}]{renema_2018_Phys.Rev.Lett.}%
  \BibitemOpen
  \bibfield  {author} {\bibinfo {author} {\bibfnamefont {J.~J.}\ \bibnamefont
  {Renema}}, \bibinfo {author} {\bibfnamefont {A.}~\bibnamefont {Menssen}},
  \bibinfo {author} {\bibfnamefont {W.~R.}\ \bibnamefont {Clements}}, \bibinfo
  {author} {\bibfnamefont {G.}~\bibnamefont {Triginer}}, \bibinfo {author}
  {\bibfnamefont {W.~S.}\ \bibnamefont {Kolthammer}},\ and\ \bibinfo {author}
  {\bibfnamefont {I.~A.}\ \bibnamefont {Walmsley}},\ }\href
  {https://doi.org/10.1103/PhysRevLett.120.220502} {\bibfield  {journal}
  {\bibinfo  {journal} {Phys. Rev. Lett.}\ }\textbf {\bibinfo {volume} {120}},\
  \bibinfo {pages} {220502} (\bibinfo {year} {2018}{\natexlab{a}})}\BibitemShut
  {NoStop}%
\bibitem [{\citenamefont
  {Shchesnovich}(2020)}]{shchesnovich_2020_Int.J.QuantumInform.}%
  \BibitemOpen
  \bibfield  {author} {\bibinfo {author} {\bibfnamefont {V.~S.}\ \bibnamefont
  {Shchesnovich}},\ }\href {https://doi.org/10.1142/S0219749920500446}
  {\bibfield  {journal} {\bibinfo  {journal} {Int. J. Quantum Inform.}\
  }\textbf {\bibinfo {volume} {18}},\ \bibinfo {pages} {2050044} (\bibinfo
  {year} {2020})}\BibitemShut {NoStop}%
\bibitem [{\citenamefont {Villalonga}\ \emph {et~al.}(2021)\citenamefont
  {Villalonga}, \citenamefont {Niu}, \citenamefont {Li}, \citenamefont {Neven},
  \citenamefont {Platt}, \citenamefont {Smelyanskiy},\ and\ \citenamefont
  {Boixo}}]{villalonga_2021_ArXiv}%
  \BibitemOpen
  \bibfield  {author} {\bibinfo {author} {\bibfnamefont {B.}~\bibnamefont
  {Villalonga}}, \bibinfo {author} {\bibfnamefont {M.~Y.}\ \bibnamefont {Niu}},
  \bibinfo {author} {\bibfnamefont {L.}~\bibnamefont {Li}}, \bibinfo {author}
  {\bibfnamefont {H.}~\bibnamefont {Neven}}, \bibinfo {author} {\bibfnamefont
  {J.~C.}\ \bibnamefont {Platt}}, \bibinfo {author} {\bibfnamefont {V.~N.}\
  \bibnamefont {Smelyanskiy}},\ and\ \bibinfo {author} {\bibfnamefont
  {S.}~\bibnamefont {Boixo}},\ }\href@noop {} {\bibfield  {journal} {\bibinfo
  {journal} {arXiv:2109.11525 [quant-ph]}\ } (\bibinfo {year} {2021})},\
  \Eprint {https://arxiv.org/abs/2109.11525} {arXiv:2109.11525 [quant-ph]}
  \BibitemShut {NoStop}%
\bibitem [{\citenamefont {Aaronson}\ and\ \citenamefont
  {Brod}(2016)}]{aaronson_2016_PhysRevAa}%
  \BibitemOpen
  \bibfield  {author} {\bibinfo {author} {\bibfnamefont {S.}~\bibnamefont
  {Aaronson}}\ and\ \bibinfo {author} {\bibfnamefont {D.~J.}\ \bibnamefont
  {Brod}},\ }\href {https://doi.org/10.1103/PhysRevA.93.012335} {\bibfield
  {journal} {\bibinfo  {journal} {Phys. Rev. A}\ }\textbf {\bibinfo {volume}
  {93}},\ \bibinfo {pages} {012335} (\bibinfo {year} {2016})}\BibitemShut
  {NoStop}%
\bibitem [{\citenamefont {Renema}\ \emph
  {et~al.}(2018{\natexlab{b}})\citenamefont {Renema}, \citenamefont
  {Shchesnovich},\ and\ \citenamefont {{Garcia-Patron}}}]{renema_2018_ArXiv}%
  \BibitemOpen
  \bibfield  {author} {\bibinfo {author} {\bibfnamefont {J.~J.}\ \bibnamefont
  {Renema}}, \bibinfo {author} {\bibfnamefont {V.}~\bibnamefont
  {Shchesnovich}},\ and\ \bibinfo {author} {\bibfnamefont {R.}~\bibnamefont
  {{Garcia-Patron}}},\ }\href@noop {} {\bibfield  {journal} {\bibinfo
  {journal} {arXiv:1809.01953 [cond-mat, physics:physics, physics:quant-ph]}\ }
  (\bibinfo {year} {2018}{\natexlab{b}})},\ \bibinfo {note} {unpublished},\
  \Eprint {https://arxiv.org/abs/1809.01953} {arXiv:1809.01953 [cond-mat,
  physics:physics, physics:quant-ph]} \BibitemShut {NoStop}%
\bibitem [{\citenamefont {{Garc{\'i}a-Patr{\'o}n}}\ \emph
  {et~al.}(2019)\citenamefont {{Garc{\'i}a-Patr{\'o}n}}, \citenamefont
  {Renema},\ and\ \citenamefont {Shchesnovich}}]{garcia-patron_2019_Quantum}%
  \BibitemOpen
  \bibfield  {author} {\bibinfo {author} {\bibfnamefont {R.}~\bibnamefont
  {{Garc{\'i}a-Patr{\'o}n}}}, \bibinfo {author} {\bibfnamefont {J.~J.}\
  \bibnamefont {Renema}},\ and\ \bibinfo {author} {\bibfnamefont
  {V.}~\bibnamefont {Shchesnovich}},\ }\href
  {https://doi.org/10.22331/q-2019-08-05-169} {\bibfield  {journal} {\bibinfo
  {journal} {Quantum}\ }\textbf {\bibinfo {volume} {3}},\ \bibinfo {pages}
  {169} (\bibinfo {year} {2019})}\BibitemShut {NoStop}%
\bibitem [{\citenamefont {Brod}\ and\ \citenamefont
  {Oszmaniec}(2020)}]{Brod2020}%
  \BibitemOpen
  \bibfield  {author} {\bibinfo {author} {\bibfnamefont {D.~J.}\ \bibnamefont
  {Brod}}\ and\ \bibinfo {author} {\bibfnamefont {M.}~\bibnamefont
  {Oszmaniec}},\ }\href {https://doi.org/10.22331/q-2020-05-14-267} {\bibfield
  {journal} {\bibinfo  {journal} {Quantum}\ }\textbf {\bibinfo {volume} {4}},\
  \bibinfo {pages} {267} (\bibinfo {year} {2020})}\BibitemShut {NoStop}%
\bibitem [{\citenamefont {Knill}\ \emph {et~al.}(2001)\citenamefont {Knill},
  \citenamefont {Laflamme},\ and\ \citenamefont {Milburn}}]{knill_2001_Nature}%
  \BibitemOpen
  \bibfield  {author} {\bibinfo {author} {\bibfnamefont {E.}~\bibnamefont
  {Knill}}, \bibinfo {author} {\bibfnamefont {R.}~\bibnamefont {Laflamme}},\
  and\ \bibinfo {author} {\bibfnamefont {G.~J.}\ \bibnamefont {Milburn}},\
  }\href {https://doi.org/10.1038/35051009} {\bibfield  {journal} {\bibinfo
  {journal} {Nature}\ }\textbf {\bibinfo {volume} {409}},\ \bibinfo {pages}
  {46} (\bibinfo {year} {2001})}\BibitemShut {NoStop}%
\bibitem [{\citenamefont {Browne}\ and\ \citenamefont
  {Rudolph}(2005)}]{browne_2005_Phys.Rev.Lett.}%
  \BibitemOpen
  \bibfield  {author} {\bibinfo {author} {\bibfnamefont {D.~E.}\ \bibnamefont
  {Browne}}\ and\ \bibinfo {author} {\bibfnamefont {T.}~\bibnamefont
  {Rudolph}},\ }\href {https://doi.org/10.1103/PhysRevLett.95.010501}
  {\bibfield  {journal} {\bibinfo  {journal} {Phys. Rev. Lett.}\ }\textbf
  {\bibinfo {volume} {95}},\ \bibinfo {pages} {010501} (\bibinfo {year}
  {2005})}\BibitemShut {NoStop}%
\bibitem [{\citenamefont {Hong}\ \emph {et~al.}(1987)\citenamefont {Hong},
  \citenamefont {Ou},\ and\ \citenamefont {Mandel}}]{Hong1987}%
  \BibitemOpen
  \bibfield  {author} {\bibinfo {author} {\bibfnamefont {C.~K.}\ \bibnamefont
  {Hong}}, \bibinfo {author} {\bibfnamefont {Z.~Y.}\ \bibnamefont {Ou}},\ and\
  \bibinfo {author} {\bibfnamefont {L.}~\bibnamefont {Mandel}},\ }\href
  {https://doi.org/10.1103/PhysRevLett.59.2044} {\bibfield  {journal} {\bibinfo
   {journal} {Phys. Rev. Lett.}\ }\textbf {\bibinfo {volume} {59}},\ \bibinfo
  {pages} {2044} (\bibinfo {year} {1987})}\BibitemShut {NoStop}%
\bibitem [{\citenamefont {Tichy}\ \emph {et~al.}(2014)\citenamefont {Tichy},
  \citenamefont {Mayer}, \citenamefont {Buchleitner},\ and\ \citenamefont
  {M{\o}lmer}}]{tichy_2014_Phys.Rev.Lett.}%
  \BibitemOpen
  \bibfield  {author} {\bibinfo {author} {\bibfnamefont {M.~C.}\ \bibnamefont
  {Tichy}}, \bibinfo {author} {\bibfnamefont {K.}~\bibnamefont {Mayer}},
  \bibinfo {author} {\bibfnamefont {A.}~\bibnamefont {Buchleitner}},\ and\
  \bibinfo {author} {\bibfnamefont {K.}~\bibnamefont {M{\o}lmer}},\ }\href
  {https://doi.org/10.1103/PhysRevLett.113.020502} {\bibfield  {journal}
  {\bibinfo  {journal} {Phys. Rev. Lett.}\ }\textbf {\bibinfo {volume} {113}},\
  \bibinfo {pages} {020502} (\bibinfo {year} {2014})}\BibitemShut {NoStop}%
\bibitem [{\citenamefont {Carolan}\ \emph {et~al.}(2014)\citenamefont
  {Carolan}, \citenamefont {Meinecke}, \citenamefont {Shadbolt}, \citenamefont
  {Russell}, \citenamefont {Ismail}, \citenamefont {W{\"o}rhoff}, \citenamefont
  {Rudolph}, \citenamefont {Thompson}, \citenamefont {O'Brien}, \citenamefont
  {Matthews},\ and\ \citenamefont {Laing}}]{carolan_2014_NaturePhoton}%
  \BibitemOpen
  \bibfield  {author} {\bibinfo {author} {\bibfnamefont {J.}~\bibnamefont
  {Carolan}}, \bibinfo {author} {\bibfnamefont {J.~D.~A.}\ \bibnamefont
  {Meinecke}}, \bibinfo {author} {\bibfnamefont {P.~J.}\ \bibnamefont
  {Shadbolt}}, \bibinfo {author} {\bibfnamefont {N.~J.}\ \bibnamefont
  {Russell}}, \bibinfo {author} {\bibfnamefont {N.}~\bibnamefont {Ismail}},
  \bibinfo {author} {\bibfnamefont {K.}~\bibnamefont {W{\"o}rhoff}}, \bibinfo
  {author} {\bibfnamefont {T.}~\bibnamefont {Rudolph}}, \bibinfo {author}
  {\bibfnamefont {M.~G.}\ \bibnamefont {Thompson}}, \bibinfo {author}
  {\bibfnamefont {J.~L.}\ \bibnamefont {O'Brien}}, \bibinfo {author}
  {\bibfnamefont {J.~C.~F.}\ \bibnamefont {Matthews}},\ and\ \bibinfo {author}
  {\bibfnamefont {A.}~\bibnamefont {Laing}},\ }\href
  {https://doi.org/10.1038/nphoton.2014.152} {\bibfield  {journal} {\bibinfo
  {journal} {Nature Photon}\ }\textbf {\bibinfo {volume} {8}},\ \bibinfo
  {pages} {621} (\bibinfo {year} {2014})}\BibitemShut {NoStop}%
\bibitem [{\citenamefont {Bentivegna}\ \emph {et~al.}(2014)\citenamefont
  {Bentivegna}, \citenamefont {Spagnolo}, \citenamefont {Vitelli},
  \citenamefont {Brod}, \citenamefont {Crespi}, \citenamefont {Flamini},
  \citenamefont {Ramponi}, \citenamefont {Mataloni}, \citenamefont {Osellame},
  \citenamefont {Galv{\~a}o},\ and\ \citenamefont
  {Sciarrino}}]{bentivegna_2014_Int.J.QuantumInform.}%
  \BibitemOpen
  \bibfield  {author} {\bibinfo {author} {\bibfnamefont {M.}~\bibnamefont
  {Bentivegna}}, \bibinfo {author} {\bibfnamefont {N.}~\bibnamefont
  {Spagnolo}}, \bibinfo {author} {\bibfnamefont {C.}~\bibnamefont {Vitelli}},
  \bibinfo {author} {\bibfnamefont {D.~J.}\ \bibnamefont {Brod}}, \bibinfo
  {author} {\bibfnamefont {A.}~\bibnamefont {Crespi}}, \bibinfo {author}
  {\bibfnamefont {F.}~\bibnamefont {Flamini}}, \bibinfo {author} {\bibfnamefont
  {R.}~\bibnamefont {Ramponi}}, \bibinfo {author} {\bibfnamefont
  {P.}~\bibnamefont {Mataloni}}, \bibinfo {author} {\bibfnamefont
  {R.}~\bibnamefont {Osellame}}, \bibinfo {author} {\bibfnamefont {E.~F.}\
  \bibnamefont {Galv{\~a}o}},\ and\ \bibinfo {author} {\bibfnamefont
  {F.}~\bibnamefont {Sciarrino}},\ }\href
  {https://doi.org/10.1142/S021974991560028X} {\bibfield  {journal} {\bibinfo
  {journal} {Int. J. Quantum Inform.}\ }\textbf {\bibinfo {volume} {12}},\
  \bibinfo {pages} {1560028} (\bibinfo {year} {2014})}\BibitemShut {NoStop}%
\bibitem [{\citenamefont {Crespi}(2015)}]{crespi_2015_Phys.Rev.A}%
  \BibitemOpen
  \bibfield  {author} {\bibinfo {author} {\bibfnamefont {A.}~\bibnamefont
  {Crespi}},\ }\href {https://doi.org/10.1103/PhysRevA.91.013811} {\bibfield
  {journal} {\bibinfo  {journal} {Phys. Rev. A}\ }\textbf {\bibinfo {volume}
  {91}},\ \bibinfo {pages} {013811} (\bibinfo {year} {2015})}\BibitemShut
  {NoStop}%
\bibitem [{\citenamefont {Bentivegna}\ \emph {et~al.}(2016)\citenamefont
  {Bentivegna}, \citenamefont {Spagnolo},\ and\ \citenamefont
  {Sciarrino}}]{bentivegna_2016_NewJ.Phys.}%
  \BibitemOpen
  \bibfield  {author} {\bibinfo {author} {\bibfnamefont {M.}~\bibnamefont
  {Bentivegna}}, \bibinfo {author} {\bibfnamefont {N.}~\bibnamefont
  {Spagnolo}},\ and\ \bibinfo {author} {\bibfnamefont {F.}~\bibnamefont
  {Sciarrino}},\ }\href {https://doi.org/10.1088/1367-2630/18/4/041001}
  {\bibfield  {journal} {\bibinfo  {journal} {New J. Phys.}\ }\textbf {\bibinfo
  {volume} {18}},\ \bibinfo {pages} {041001} (\bibinfo {year}
  {2016})}\BibitemShut {NoStop}%
\bibitem [{\citenamefont {Walschaers}\ \emph {et~al.}(2016)\citenamefont
  {Walschaers}, \citenamefont {Kuipers}, \citenamefont {Urbina}, \citenamefont
  {Mayer}, \citenamefont {Tichy}, \citenamefont {Richter},\ and\ \citenamefont
  {Buchleitner}}]{Walschaers2016}%
  \BibitemOpen
  \bibfield  {author} {\bibinfo {author} {\bibfnamefont {M.}~\bibnamefont
  {Walschaers}}, \bibinfo {author} {\bibfnamefont {J.}~\bibnamefont {Kuipers}},
  \bibinfo {author} {\bibfnamefont {J.-D.}\ \bibnamefont {Urbina}}, \bibinfo
  {author} {\bibfnamefont {K.}~\bibnamefont {Mayer}}, \bibinfo {author}
  {\bibfnamefont {M.~C.}\ \bibnamefont {Tichy}}, \bibinfo {author}
  {\bibfnamefont {K.}~\bibnamefont {Richter}},\ and\ \bibinfo {author}
  {\bibfnamefont {A.}~\bibnamefont {Buchleitner}},\ }\href
  {https://doi.org/10.1088/1367-2630/18/3/032001} {\bibfield  {journal}
  {\bibinfo  {journal} {New J. Phys.}\ }\textbf {\bibinfo {volume} {18}},\
  \bibinfo {pages} {032001} (\bibinfo {year} {2016})}\BibitemShut {NoStop}%
\bibitem [{\citenamefont {Dittel}\ \emph {et~al.}(2017)\citenamefont {Dittel},
  \citenamefont {Keil},\ and\ \citenamefont
  {Weihs}}]{dittel_2017_QuantumSci.Technol.}%
  \BibitemOpen
  \bibfield  {author} {\bibinfo {author} {\bibfnamefont {C.}~\bibnamefont
  {Dittel}}, \bibinfo {author} {\bibfnamefont {R.}~\bibnamefont {Keil}},\ and\
  \bibinfo {author} {\bibfnamefont {G.}~\bibnamefont {Weihs}},\ }\href
  {https://doi.org/10.1088/2058-9565/aa540c} {\bibfield  {journal} {\bibinfo
  {journal} {Quantum Sci. Technol.}\ }\textbf {\bibinfo {volume} {2}},\
  \bibinfo {pages} {015003} (\bibinfo {year} {2017})}\BibitemShut {NoStop}%
\bibitem [{\citenamefont {Viggianiello}\ \emph
  {et~al.}(2018{\natexlab{a}})\citenamefont {Viggianiello}, \citenamefont
  {Flamini}, \citenamefont {Bentivegna}, \citenamefont {Spagnolo},
  \citenamefont {Crespi}, \citenamefont {Brod}, \citenamefont {Galv{\~a}o},
  \citenamefont {Osellame},\ and\ \citenamefont
  {Sciarrino}}]{viggianiello_2018_ScienceBulletin}%
  \BibitemOpen
  \bibfield  {author} {\bibinfo {author} {\bibfnamefont {N.}~\bibnamefont
  {Viggianiello}}, \bibinfo {author} {\bibfnamefont {F.}~\bibnamefont
  {Flamini}}, \bibinfo {author} {\bibfnamefont {M.}~\bibnamefont {Bentivegna}},
  \bibinfo {author} {\bibfnamefont {N.}~\bibnamefont {Spagnolo}}, \bibinfo
  {author} {\bibfnamefont {A.}~\bibnamefont {Crespi}}, \bibinfo {author}
  {\bibfnamefont {D.~J.}\ \bibnamefont {Brod}}, \bibinfo {author}
  {\bibfnamefont {E.~F.}\ \bibnamefont {Galv{\~a}o}}, \bibinfo {author}
  {\bibfnamefont {R.}~\bibnamefont {Osellame}},\ and\ \bibinfo {author}
  {\bibfnamefont {F.}~\bibnamefont {Sciarrino}},\ }\href
  {https://doi.org/10.1016/j.scib.2018.10.009} {\bibfield  {journal} {\bibinfo
  {journal} {Science Bulletin}\ }\textbf {\bibinfo {volume} {63}},\ \bibinfo
  {pages} {1470} (\bibinfo {year} {2018}{\natexlab{a}})}\BibitemShut {NoStop}%
\bibitem [{\citenamefont {Giordani}\ \emph {et~al.}(2018)\citenamefont
  {Giordani}, \citenamefont {Flamini}, \citenamefont {Pompili}, \citenamefont
  {Viggianiello}, \citenamefont {Spagnolo}, \citenamefont {Crespi},
  \citenamefont {Osellame}, \citenamefont {Wiebe}, \citenamefont {Walschaers},
  \citenamefont {Buchleitner},\ and\ \citenamefont
  {Sciarrino}}]{giordani_2018_NaturePhoton}%
  \BibitemOpen
  \bibfield  {author} {\bibinfo {author} {\bibfnamefont {T.}~\bibnamefont
  {Giordani}}, \bibinfo {author} {\bibfnamefont {F.}~\bibnamefont {Flamini}},
  \bibinfo {author} {\bibfnamefont {M.}~\bibnamefont {Pompili}}, \bibinfo
  {author} {\bibfnamefont {N.}~\bibnamefont {Viggianiello}}, \bibinfo {author}
  {\bibfnamefont {N.}~\bibnamefont {Spagnolo}}, \bibinfo {author}
  {\bibfnamefont {A.}~\bibnamefont {Crespi}}, \bibinfo {author} {\bibfnamefont
  {R.}~\bibnamefont {Osellame}}, \bibinfo {author} {\bibfnamefont
  {N.}~\bibnamefont {Wiebe}}, \bibinfo {author} {\bibfnamefont
  {M.}~\bibnamefont {Walschaers}}, \bibinfo {author} {\bibfnamefont
  {A.}~\bibnamefont {Buchleitner}},\ and\ \bibinfo {author} {\bibfnamefont
  {F.}~\bibnamefont {Sciarrino}},\ }\href
  {https://doi.org/10.1038/s41566-018-0097-4} {\bibfield  {journal} {\bibinfo
  {journal} {Nature Photon}\ }\textbf {\bibinfo {volume} {12}},\ \bibinfo
  {pages} {173} (\bibinfo {year} {2018})}\BibitemShut {NoStop}%
\bibitem [{\citenamefont {Viggianiello}\ \emph
  {et~al.}(2018{\natexlab{b}})\citenamefont {Viggianiello}, \citenamefont
  {Flamini}, \citenamefont {Innocenti}, \citenamefont {Cozzolino},
  \citenamefont {Bentivegna}, \citenamefont {Spagnolo}, \citenamefont {Crespi},
  \citenamefont {Brod}, \citenamefont {Galv{\~a}o}, \citenamefont {Osellame},\
  and\ \citenamefont {Sciarrino}}]{viggianiello_2018_NewJ.Phys.}%
  \BibitemOpen
  \bibfield  {author} {\bibinfo {author} {\bibfnamefont {N.}~\bibnamefont
  {Viggianiello}}, \bibinfo {author} {\bibfnamefont {F.}~\bibnamefont
  {Flamini}}, \bibinfo {author} {\bibfnamefont {L.}~\bibnamefont {Innocenti}},
  \bibinfo {author} {\bibfnamefont {D.}~\bibnamefont {Cozzolino}}, \bibinfo
  {author} {\bibfnamefont {M.}~\bibnamefont {Bentivegna}}, \bibinfo {author}
  {\bibfnamefont {N.}~\bibnamefont {Spagnolo}}, \bibinfo {author}
  {\bibfnamefont {A.}~\bibnamefont {Crespi}}, \bibinfo {author} {\bibfnamefont
  {D.~J.}\ \bibnamefont {Brod}}, \bibinfo {author} {\bibfnamefont {E.~F.}\
  \bibnamefont {Galv{\~a}o}}, \bibinfo {author} {\bibfnamefont
  {R.}~\bibnamefont {Osellame}},\ and\ \bibinfo {author} {\bibfnamefont
  {F.}~\bibnamefont {Sciarrino}},\ }\href
  {https://doi.org/10.1088/1367-2630/aaad92} {\bibfield  {journal} {\bibinfo
  {journal} {New J. Phys.}\ }\textbf {\bibinfo {volume} {20}},\ \bibinfo
  {pages} {033017} (\bibinfo {year} {2018}{\natexlab{b}})}\BibitemShut
  {NoStop}%
\bibitem [{\citenamefont {Agresti}\ \emph {et~al.}(2019)\citenamefont
  {Agresti}, \citenamefont {Viggianiello}, \citenamefont {Flamini},
  \citenamefont {Spagnolo}, \citenamefont {Crespi}, \citenamefont {Osellame},
  \citenamefont {Wiebe},\ and\ \citenamefont
  {Sciarrino}}]{agresti_2019_Phys.Rev.X}%
  \BibitemOpen
  \bibfield  {author} {\bibinfo {author} {\bibfnamefont {I.}~\bibnamefont
  {Agresti}}, \bibinfo {author} {\bibfnamefont {N.}~\bibnamefont
  {Viggianiello}}, \bibinfo {author} {\bibfnamefont {F.}~\bibnamefont
  {Flamini}}, \bibinfo {author} {\bibfnamefont {N.}~\bibnamefont {Spagnolo}},
  \bibinfo {author} {\bibfnamefont {A.}~\bibnamefont {Crespi}}, \bibinfo
  {author} {\bibfnamefont {R.}~\bibnamefont {Osellame}}, \bibinfo {author}
  {\bibfnamefont {N.}~\bibnamefont {Wiebe}},\ and\ \bibinfo {author}
  {\bibfnamefont {F.}~\bibnamefont {Sciarrino}},\ }\href
  {https://doi.org/10.1103/PhysRevX.9.011013} {\bibfield  {journal} {\bibinfo
  {journal} {Phys. Rev. X}\ }\textbf {\bibinfo {volume} {9}},\ \bibinfo {pages}
  {011013} (\bibinfo {year} {2019})}\BibitemShut {NoStop}%
\bibitem [{\citenamefont {Brod}\ \emph {et~al.}(2019)\citenamefont {Brod},
  \citenamefont {Galv{\~a}o}, \citenamefont {Viggianiello}, \citenamefont
  {Flamini}, \citenamefont {Spagnolo},\ and\ \citenamefont
  {Sciarrino}}]{brod_2019_Phys.Rev.Lett.}%
  \BibitemOpen
  \bibfield  {author} {\bibinfo {author} {\bibfnamefont {D.~J.}\ \bibnamefont
  {Brod}}, \bibinfo {author} {\bibfnamefont {E.~F.}\ \bibnamefont
  {Galv{\~a}o}}, \bibinfo {author} {\bibfnamefont {N.}~\bibnamefont
  {Viggianiello}}, \bibinfo {author} {\bibfnamefont {F.}~\bibnamefont
  {Flamini}}, \bibinfo {author} {\bibfnamefont {N.}~\bibnamefont {Spagnolo}},\
  and\ \bibinfo {author} {\bibfnamefont {F.}~\bibnamefont {Sciarrino}},\ }\href
  {https://doi.org/10.1103/PhysRevLett.122.063602} {\bibfield  {journal}
  {\bibinfo  {journal} {Phys. Rev. Lett.}\ }\textbf {\bibinfo {volume} {122}},\
  \bibinfo {pages} {063602} (\bibinfo {year} {2019})}\BibitemShut {NoStop}%
\bibitem [{\citenamefont {Terhal}(2000)}]{Terhal2000}%
  \BibitemOpen
  \bibfield  {author} {\bibinfo {author} {\bibfnamefont {B.~M.}\ \bibnamefont
  {Terhal}},\ }\href
  {https://doi.org/https://doi.org/10.1016/S0375-9601(00)00401-1} {\bibfield
  {journal} {\bibinfo  {journal} {Physics Letters A}\ }\textbf {\bibinfo
  {volume} {271}},\ \bibinfo {pages} {319} (\bibinfo {year}
  {2000})}\BibitemShut {NoStop}%
\bibitem [{\citenamefont {Shchesnovich}(2016)}]{Shchesnovich2016}%
  \BibitemOpen
  \bibfield  {author} {\bibinfo {author} {\bibfnamefont {V.~S.}\ \bibnamefont
  {Shchesnovich}},\ }\href {https://doi.org/10.1103/PhysRevLett.116.123601}
  {\bibfield  {journal} {\bibinfo  {journal} {Phys. Rev. Lett.}\ }\textbf
  {\bibinfo {volume} {116}},\ \bibinfo {pages} {123601} (\bibinfo {year}
  {2016})}\BibitemShut {NoStop}%
\bibitem [{\citenamefont {Gerhardt}\ \emph {et~al.}(2011)\citenamefont
  {Gerhardt}, \citenamefont {Liu}, \citenamefont {Lamas-Linares}, \citenamefont
  {Skaar}, \citenamefont {Scarani}, \citenamefont {Makarov},\ and\
  \citenamefont {Kurtsiefer}}]{Gerhardt2011}%
  \BibitemOpen
  \bibfield  {author} {\bibinfo {author} {\bibfnamefont {I.}~\bibnamefont
  {Gerhardt}}, \bibinfo {author} {\bibfnamefont {Q.}~\bibnamefont {Liu}},
  \bibinfo {author} {\bibfnamefont {A.}~\bibnamefont {Lamas-Linares}}, \bibinfo
  {author} {\bibfnamefont {J.}~\bibnamefont {Skaar}}, \bibinfo {author}
  {\bibfnamefont {V.}~\bibnamefont {Scarani}}, \bibinfo {author} {\bibfnamefont
  {V.}~\bibnamefont {Makarov}},\ and\ \bibinfo {author} {\bibfnamefont
  {C.}~\bibnamefont {Kurtsiefer}},\ }\href
  {https://doi.org/10.1103/PhysRevLett.107.170404} {\bibfield  {journal}
  {\bibinfo  {journal} {Phys. Rev. Lett.}\ }\textbf {\bibinfo {volume} {107}},\
  \bibinfo {pages} {170404} (\bibinfo {year} {2011})}\BibitemShut {NoStop}%
\bibitem [{\citenamefont {Eisert}\ \emph {et~al.}(2020)\citenamefont {Eisert},
  \citenamefont {Hangleiter}, \citenamefont {Walk}, \citenamefont {Roth},
  \citenamefont {Markham}, \citenamefont {Parekh}, \citenamefont {Chabaud},\
  and\ \citenamefont {Kashefi}}]{BenchmarkingReview}%
  \BibitemOpen
  \bibfield  {author} {\bibinfo {author} {\bibfnamefont {J.}~\bibnamefont
  {Eisert}}, \bibinfo {author} {\bibfnamefont {D.}~\bibnamefont {Hangleiter}},
  \bibinfo {author} {\bibfnamefont {N.}~\bibnamefont {Walk}}, \bibinfo {author}
  {\bibfnamefont {I.}~\bibnamefont {Roth}}, \bibinfo {author} {\bibfnamefont
  {D.}~\bibnamefont {Markham}}, \bibinfo {author} {\bibfnamefont
  {R.}~\bibnamefont {Parekh}}, \bibinfo {author} {\bibfnamefont
  {U.}~\bibnamefont {Chabaud}},\ and\ \bibinfo {author} {\bibfnamefont
  {E.}~\bibnamefont {Kashefi}},\ }\href
  {https://doi.org/10.1038/s42254-020-0186-4} {\bibfield  {journal} {\bibinfo
  {journal} {Nature Rev. Phys.}\ }\textbf {\bibinfo {volume} {2}},\ \bibinfo
  {pages} {382} (\bibinfo {year} {2020})}\BibitemShut {NoStop}%
\bibitem [{\citenamefont {Pironio}\ \emph {et~al.}(2010)\citenamefont
  {Pironio}, \citenamefont {Ac{\'\i}n}, \citenamefont {Massar}, \citenamefont
  {de~La~Giroday}, \citenamefont {Matsukevich}, \citenamefont {Maunz},
  \citenamefont {Olmschenk}, \citenamefont {Hayes}, \citenamefont {Luo},
  \citenamefont {Manning} \emph {et~al.}}]{pironio2010random}%
  \BibitemOpen
  \bibfield  {author} {\bibinfo {author} {\bibfnamefont {S.}~\bibnamefont
  {Pironio}}, \bibinfo {author} {\bibfnamefont {A.}~\bibnamefont {Ac{\'\i}n}},
  \bibinfo {author} {\bibfnamefont {S.}~\bibnamefont {Massar}}, \bibinfo
  {author} {\bibfnamefont {A.~B.}\ \bibnamefont {de~La~Giroday}}, \bibinfo
  {author} {\bibfnamefont {D.~N.}\ \bibnamefont {Matsukevich}}, \bibinfo
  {author} {\bibfnamefont {P.}~\bibnamefont {Maunz}}, \bibinfo {author}
  {\bibfnamefont {S.}~\bibnamefont {Olmschenk}}, \bibinfo {author}
  {\bibfnamefont {D.}~\bibnamefont {Hayes}}, \bibinfo {author} {\bibfnamefont
  {L.}~\bibnamefont {Luo}}, \bibinfo {author} {\bibfnamefont {T.~A.}\
  \bibnamefont {Manning}}, \emph {et~al.},\ }\href@noop {} {\bibfield
  {journal} {\bibinfo  {journal} {Nature}\ }\textbf {\bibinfo {volume} {464}},\
  \bibinfo {pages} {1021} (\bibinfo {year} {2010})}\BibitemShut {NoStop}%
\bibitem [{\citenamefont {Ac{\'\i}n}\ and\ \citenamefont
  {Masanes}(2016)}]{Acin:2016ke}%
  \BibitemOpen
  \bibfield  {author} {\bibinfo {author} {\bibfnamefont {A.}~\bibnamefont
  {Ac{\'\i}n}}\ and\ \bibinfo {author} {\bibfnamefont {L.}~\bibnamefont
  {Masanes}},\ }\href@noop {} {\bibfield  {journal} {\bibinfo  {journal}
  {Nature}\ }\textbf {\bibinfo {volume} {540}},\ \bibinfo {pages} {213}
  (\bibinfo {year} {2016})}\BibitemShut {NoStop}%
\bibitem [{\citenamefont {Brunner}\ \emph {et~al.}(2014)\citenamefont
  {Brunner}, \citenamefont {Cavalcanti}, \citenamefont {Pironio}, \citenamefont
  {Scarani},\ and\ \citenamefont {Wehner}}]{Brunner:2014vw}%
  \BibitemOpen
  \bibfield  {author} {\bibinfo {author} {\bibfnamefont {N.}~\bibnamefont
  {Brunner}}, \bibinfo {author} {\bibfnamefont {D.}~\bibnamefont {Cavalcanti}},
  \bibinfo {author} {\bibfnamefont {S.}~\bibnamefont {Pironio}}, \bibinfo
  {author} {\bibfnamefont {V.}~\bibnamefont {Scarani}},\ and\ \bibinfo {author}
  {\bibfnamefont {S.}~\bibnamefont {Wehner}},\ }\href@noop {} {\bibfield
  {journal} {\bibinfo  {journal} {Reviews of Modern Physics}\ }\textbf
  {\bibinfo {volume} {86}},\ \bibinfo {pages} {419} (\bibinfo {year}
  {2014})}\BibitemShut {NoStop}%
\bibitem [{\citenamefont {Liu}\ \emph {et~al.}(2021)\citenamefont {Liu},
  \citenamefont {Li}, \citenamefont {Ragy}, \citenamefont {Zhao}, \citenamefont
  {Bai}, \citenamefont {Liu}, \citenamefont {Brown}, \citenamefont {Zhang},
  \citenamefont {Colbeck}, \citenamefont {Fan}, \citenamefont {Zhang},\ and\
  \citenamefont {Pan}}]{Liu:2021ba}%
  \BibitemOpen
  \bibfield  {author} {\bibinfo {author} {\bibfnamefont {W.-Z.}\ \bibnamefont
  {Liu}}, \bibinfo {author} {\bibfnamefont {M.-H.}\ \bibnamefont {Li}},
  \bibinfo {author} {\bibfnamefont {S.}~\bibnamefont {Ragy}}, \bibinfo {author}
  {\bibfnamefont {S.-R.}\ \bibnamefont {Zhao}}, \bibinfo {author}
  {\bibfnamefont {B.}~\bibnamefont {Bai}}, \bibinfo {author} {\bibfnamefont
  {Y.}~\bibnamefont {Liu}}, \bibinfo {author} {\bibfnamefont {P.~J.}\
  \bibnamefont {Brown}}, \bibinfo {author} {\bibfnamefont {J.}~\bibnamefont
  {Zhang}}, \bibinfo {author} {\bibfnamefont {R.}~\bibnamefont {Colbeck}},
  \bibinfo {author} {\bibfnamefont {J.}~\bibnamefont {Fan}}, \bibinfo {author}
  {\bibfnamefont {Q.}~\bibnamefont {Zhang}},\ and\ \bibinfo {author}
  {\bibfnamefont {J.-W.}\ \bibnamefont {Pan}},\ }\href@noop {} {\bibfield
  {journal} {\bibinfo  {journal} {Nature Physics}\ }\textbf {\bibinfo {volume}
  {17}},\ \bibinfo {pages} {448} (\bibinfo {year} {2021})}\BibitemShut
  {NoStop}%
\bibitem [{\citenamefont {Shalm}\ \emph {et~al.}(2021)\citenamefont {Shalm},
  \citenamefont {Zhang}, \citenamefont {Bienfang}, \citenamefont {Schlager},
  \citenamefont {Stevens}, \citenamefont {Mazurek}, \citenamefont {Abellan},
  \citenamefont {Amaya}, \citenamefont {Mitchell}, \citenamefont {Alhejji},
  \citenamefont {Fu}, \citenamefont {Ornstein}, \citenamefont {Mirin},
  \citenamefont {Nam},\ and\ \citenamefont {Knill}}]{Shalm:2021jq}%
  \BibitemOpen
  \bibfield  {author} {\bibinfo {author} {\bibfnamefont {L.~K.}\ \bibnamefont
  {Shalm}}, \bibinfo {author} {\bibfnamefont {Y.}~\bibnamefont {Zhang}},
  \bibinfo {author} {\bibfnamefont {J.~C.}\ \bibnamefont {Bienfang}}, \bibinfo
  {author} {\bibfnamefont {C.}~\bibnamefont {Schlager}}, \bibinfo {author}
  {\bibfnamefont {M.~J.}\ \bibnamefont {Stevens}}, \bibinfo {author}
  {\bibfnamefont {M.~D.}\ \bibnamefont {Mazurek}}, \bibinfo {author}
  {\bibfnamefont {C.}~\bibnamefont {Abellan}}, \bibinfo {author} {\bibfnamefont
  {W.}~\bibnamefont {Amaya}}, \bibinfo {author} {\bibfnamefont {M.~W.}\
  \bibnamefont {Mitchell}}, \bibinfo {author} {\bibfnamefont {M.~A.}\
  \bibnamefont {Alhejji}}, \bibinfo {author} {\bibfnamefont {H.}~\bibnamefont
  {Fu}}, \bibinfo {author} {\bibfnamefont {J.}~\bibnamefont {Ornstein}},
  \bibinfo {author} {\bibfnamefont {R.~P.}\ \bibnamefont {Mirin}}, \bibinfo
  {author} {\bibfnamefont {S.~W.}\ \bibnamefont {Nam}},\ and\ \bibinfo {author}
  {\bibfnamefont {E.}~\bibnamefont {Knill}},\ }\href@noop {} {\bibfield
  {journal} {\bibinfo  {journal} {Nature Physics}\ }\textbf {\bibinfo {volume}
  {17}},\ \bibinfo {pages} {452} (\bibinfo {year} {2021})}\BibitemShut
  {NoStop}%
\bibitem [{\citenamefont {Paw{\l}owski}\ and\ \citenamefont
  {Brunner}(2011)}]{pawlowski2011semi}%
  \BibitemOpen
  \bibfield  {author} {\bibinfo {author} {\bibfnamefont {M.}~\bibnamefont
  {Paw{\l}owski}}\ and\ \bibinfo {author} {\bibfnamefont {N.}~\bibnamefont
  {Brunner}},\ }\href@noop {} {\bibfield  {journal} {\bibinfo  {journal}
  {Physical Review A}\ }\textbf {\bibinfo {volume} {84}},\ \bibinfo {pages}
  {010302} (\bibinfo {year} {2011})}\BibitemShut {NoStop}%
\bibitem [{\citenamefont {Cao}\ \emph {et~al.}(2016)\citenamefont {Cao},
  \citenamefont {Zhou}, \citenamefont {Yuan},\ and\ \citenamefont
  {Ma}}]{cao2016source}%
  \BibitemOpen
  \bibfield  {author} {\bibinfo {author} {\bibfnamefont {Z.}~\bibnamefont
  {Cao}}, \bibinfo {author} {\bibfnamefont {H.}~\bibnamefont {Zhou}}, \bibinfo
  {author} {\bibfnamefont {X.}~\bibnamefont {Yuan}},\ and\ \bibinfo {author}
  {\bibfnamefont {X.}~\bibnamefont {Ma}},\ }\href@noop {} {\bibfield  {journal}
  {\bibinfo  {journal} {Physical Review X}\ }\textbf {\bibinfo {volume} {6}},\
  \bibinfo {pages} {011020} (\bibinfo {year} {2016})}\BibitemShut {NoStop}%
\bibitem [{\citenamefont {Marangon}\ \emph {et~al.}(2017)\citenamefont
  {Marangon}, \citenamefont {Vallone},\ and\ \citenamefont
  {Villoresi}}]{marangon2017source}%
  \BibitemOpen
  \bibfield  {author} {\bibinfo {author} {\bibfnamefont {D.~G.}\ \bibnamefont
  {Marangon}}, \bibinfo {author} {\bibfnamefont {G.}~\bibnamefont {Vallone}},\
  and\ \bibinfo {author} {\bibfnamefont {P.}~\bibnamefont {Villoresi}},\
  }\href@noop {} {\bibfield  {journal} {\bibinfo  {journal} {Physical review
  letters}\ }\textbf {\bibinfo {volume} {118}},\ \bibinfo {pages} {060503}
  (\bibinfo {year} {2017})}\BibitemShut {NoStop}%
\bibitem [{\citenamefont {Drahi}\ \emph {et~al.}(2020)\citenamefont {Drahi},
  \citenamefont {Walk}, \citenamefont {Hoban}, \citenamefont {Fedorov},
  \citenamefont {Shakhovoy}, \citenamefont {Feimov}, \citenamefont {Kurochkin},
  \citenamefont {Kolthammer}, \citenamefont {Nunn}, \citenamefont {Barrett},\
  and\ \citenamefont {Walmsley}}]{Drahi:2020gw}%
  \BibitemOpen
  \bibfield  {author} {\bibinfo {author} {\bibfnamefont {D.}~\bibnamefont
  {Drahi}}, \bibinfo {author} {\bibfnamefont {N.}~\bibnamefont {Walk}},
  \bibinfo {author} {\bibfnamefont {M.~J.}\ \bibnamefont {Hoban}}, \bibinfo
  {author} {\bibfnamefont {A.~K.}\ \bibnamefont {Fedorov}}, \bibinfo {author}
  {\bibfnamefont {R.}~\bibnamefont {Shakhovoy}}, \bibinfo {author}
  {\bibfnamefont {A.}~\bibnamefont {Feimov}}, \bibinfo {author} {\bibfnamefont
  {Y.}~\bibnamefont {Kurochkin}}, \bibinfo {author} {\bibfnamefont {W.~S.}\
  \bibnamefont {Kolthammer}}, \bibinfo {author} {\bibfnamefont
  {J.}~\bibnamefont {Nunn}}, \bibinfo {author} {\bibfnamefont {J.}~\bibnamefont
  {Barrett}},\ and\ \bibinfo {author} {\bibfnamefont {I.~A.}\ \bibnamefont
  {Walmsley}},\ }\href@noop {} {\bibfield  {journal} {\bibinfo  {journal}
  {Physical Review X}\ }\textbf {\bibinfo {volume} {10}},\ \bibinfo {pages}
  {041048} (\bibinfo {year} {2020})}\BibitemShut {NoStop}%
\bibitem [{\citenamefont {Chaturvedi}\ and\ \citenamefont
  {Banik}(2015)}]{chaturvedi2015measurement}%
  \BibitemOpen
  \bibfield  {author} {\bibinfo {author} {\bibfnamefont {A.}~\bibnamefont
  {Chaturvedi}}\ and\ \bibinfo {author} {\bibfnamefont {M.}~\bibnamefont
  {Banik}},\ }\href@noop {} {\bibfield  {journal} {\bibinfo  {journal} {EPL
  (Europhysics Letters)}\ }\textbf {\bibinfo {volume} {112}},\ \bibinfo {pages}
  {30003} (\bibinfo {year} {2015})}\BibitemShut {NoStop}%
\bibitem [{\citenamefont {Cao}\ \emph {et~al.}(2015)\citenamefont {Cao},
  \citenamefont {Zhou},\ and\ \citenamefont {Ma}}]{cao2015loss}%
  \BibitemOpen
  \bibfield  {author} {\bibinfo {author} {\bibfnamefont {Z.}~\bibnamefont
  {Cao}}, \bibinfo {author} {\bibfnamefont {H.}~\bibnamefont {Zhou}},\ and\
  \bibinfo {author} {\bibfnamefont {X.}~\bibnamefont {Ma}},\ }\href@noop {}
  {\bibfield  {journal} {\bibinfo  {journal} {New Journal of Physics}\ }\textbf
  {\bibinfo {volume} {17}},\ \bibinfo {pages} {125011} (\bibinfo {year}
  {2015})}\BibitemShut {NoStop}%
\bibitem [{\citenamefont {Himbeeck}\ \emph {et~al.}(2017)\citenamefont
  {Himbeeck}, \citenamefont {Woodhead}, \citenamefont {Cerf}, \citenamefont
  {Garc{\'{i}}a-Patr{\'{o}}n},\ and\ \citenamefont
  {Pironio}}]{Himbeeck2017semidevice}%
  \BibitemOpen
  \bibfield  {author} {\bibinfo {author} {\bibfnamefont {T.~V.}\ \bibnamefont
  {Himbeeck}}, \bibinfo {author} {\bibfnamefont {E.}~\bibnamefont {Woodhead}},
  \bibinfo {author} {\bibfnamefont {N.~J.}\ \bibnamefont {Cerf}}, \bibinfo
  {author} {\bibfnamefont {R.}~\bibnamefont {Garc{\'{i}}a-Patr{\'{o}}n}},\ and\
  \bibinfo {author} {\bibfnamefont {S.}~\bibnamefont {Pironio}},\ }\href@noop
  {} {\bibfield  {journal} {\bibinfo  {journal} {{Quantum}}\ }\textbf {\bibinfo
  {volume} {1}},\ \bibinfo {pages} {33} (\bibinfo {year} {2017})}\BibitemShut
  {NoStop}%
\bibitem [{\citenamefont {Brask}\ \emph {et~al.}(2017)\citenamefont {Brask},
  \citenamefont {Martin}, \citenamefont {Esposito}, \citenamefont {Houlmann},
  \citenamefont {Bowles}, \citenamefont {Zbinden},\ and\ \citenamefont
  {Brunner}}]{brask2017megahertz}%
  \BibitemOpen
  \bibfield  {author} {\bibinfo {author} {\bibfnamefont {J.~B.}\ \bibnamefont
  {Brask}}, \bibinfo {author} {\bibfnamefont {A.}~\bibnamefont {Martin}},
  \bibinfo {author} {\bibfnamefont {W.}~\bibnamefont {Esposito}}, \bibinfo
  {author} {\bibfnamefont {R.}~\bibnamefont {Houlmann}}, \bibinfo {author}
  {\bibfnamefont {J.}~\bibnamefont {Bowles}}, \bibinfo {author} {\bibfnamefont
  {H.}~\bibnamefont {Zbinden}},\ and\ \bibinfo {author} {\bibfnamefont
  {N.}~\bibnamefont {Brunner}},\ }\href@noop {} {\bibfield  {journal} {\bibinfo
   {journal} {Physical Review Applied}\ }\textbf {\bibinfo {volume} {7}},\
  \bibinfo {pages} {054018} (\bibinfo {year} {2017})}\BibitemShut {NoStop}%
\bibitem [{\citenamefont {M{\'a}ttar}\ \emph {et~al.}(2017)\citenamefont
  {M{\'a}ttar}, \citenamefont {Skrzypczyk}, \citenamefont {Aguilar},
  \citenamefont {Nery}, \citenamefont {Ribeiro}, \citenamefont {Walborn},\ and\
  \citenamefont {Cavalcanti}}]{mattar2017experimental}%
  \BibitemOpen
  \bibfield  {author} {\bibinfo {author} {\bibfnamefont {A.}~\bibnamefont
  {M{\'a}ttar}}, \bibinfo {author} {\bibfnamefont {P.}~\bibnamefont
  {Skrzypczyk}}, \bibinfo {author} {\bibfnamefont {G.}~\bibnamefont {Aguilar}},
  \bibinfo {author} {\bibfnamefont {R.}~\bibnamefont {Nery}}, \bibinfo {author}
  {\bibfnamefont {P.~S.}\ \bibnamefont {Ribeiro}}, \bibinfo {author}
  {\bibfnamefont {S.}~\bibnamefont {Walborn}},\ and\ \bibinfo {author}
  {\bibfnamefont {D.}~\bibnamefont {Cavalcanti}},\ }\href@noop {} {\bibfield
  {journal} {\bibinfo  {journal} {Quantum Science and Technology}\ }\textbf
  {\bibinfo {volume} {2}},\ \bibinfo {pages} {015011} (\bibinfo {year}
  {2017})}\BibitemShut {NoStop}%
\bibitem [{\citenamefont {Kocsis}\ \emph {et~al.}(2015)\citenamefont {Kocsis},
  \citenamefont {Hall}, \citenamefont {Bennet}, \citenamefont {Saunders},\ and\
  \citenamefont {Pryde}}]{Kocsis:2015gu}%
  \BibitemOpen
  \bibfield  {author} {\bibinfo {author} {\bibfnamefont {S.}~\bibnamefont
  {Kocsis}}, \bibinfo {author} {\bibfnamefont {M.~J.~W.}\ \bibnamefont {Hall}},
  \bibinfo {author} {\bibfnamefont {A.~J.}\ \bibnamefont {Bennet}}, \bibinfo
  {author} {\bibfnamefont {D.~J.}\ \bibnamefont {Saunders}},\ and\ \bibinfo
  {author} {\bibfnamefont {G.~J.}\ \bibnamefont {Pryde}},\ }\href@noop {}
  {\bibfield  {journal} {\bibinfo  {journal} {Nature Communications}\ }\textbf
  {\bibinfo {volume} {6}},\ \bibinfo {pages} {5886} (\bibinfo {year}
  {2015})}\BibitemShut {NoStop}%
\bibitem [{\citenamefont {Carolan}\ \emph {et~al.}(2015)\citenamefont
  {Carolan}, \citenamefont {Harrold}, \citenamefont {Sparrow}, \citenamefont
  {Martín-López}, \citenamefont {Russell}, \citenamefont {Silverstone},
  \citenamefont {Shadbolt}, \citenamefont {Matsuda}, \citenamefont {Oguma},
  \citenamefont {Itoh}, \citenamefont {Marshall}, \citenamefont {Thompson},
  \citenamefont {Matthews}, \citenamefont {Hashimoto}, \citenamefont
  {O’Brien},\ and\ \citenamefont {Laing}}]{Carolan2015}%
  \BibitemOpen
  \bibfield  {author} {\bibinfo {author} {\bibfnamefont {J.}~\bibnamefont
  {Carolan}}, \bibinfo {author} {\bibfnamefont {C.}~\bibnamefont {Harrold}},
  \bibinfo {author} {\bibfnamefont {C.}~\bibnamefont {Sparrow}}, \bibinfo
  {author} {\bibfnamefont {E.}~\bibnamefont {Martín-López}}, \bibinfo
  {author} {\bibfnamefont {N.~J.}\ \bibnamefont {Russell}}, \bibinfo {author}
  {\bibfnamefont {J.~W.}\ \bibnamefont {Silverstone}}, \bibinfo {author}
  {\bibfnamefont {P.~J.}\ \bibnamefont {Shadbolt}}, \bibinfo {author}
  {\bibfnamefont {N.}~\bibnamefont {Matsuda}}, \bibinfo {author} {\bibfnamefont
  {M.}~\bibnamefont {Oguma}}, \bibinfo {author} {\bibfnamefont
  {M.}~\bibnamefont {Itoh}}, \bibinfo {author} {\bibfnamefont {G.~D.}\
  \bibnamefont {Marshall}}, \bibinfo {author} {\bibfnamefont {M.~G.}\
  \bibnamefont {Thompson}}, \bibinfo {author} {\bibfnamefont {J.~C.~F.}\
  \bibnamefont {Matthews}}, \bibinfo {author} {\bibfnamefont {T.}~\bibnamefont
  {Hashimoto}}, \bibinfo {author} {\bibfnamefont {J.~L.}\ \bibnamefont
  {O’Brien}},\ and\ \bibinfo {author} {\bibfnamefont {A.}~\bibnamefont
  {Laing}},\ }\href {https://doi.org/10.1126/science.aab3642} {\bibfield
  {journal} {\bibinfo  {journal} {Science}\ }\textbf {\bibinfo {volume}
  {349}},\ \bibinfo {pages} {711} (\bibinfo {year} {2015})},\ \Eprint
  {https://arxiv.org/abs/https://www.science.org/doi/pdf/10.1126/science.aab3642}
  {https://www.science.org/doi/pdf/10.1126/science.aab3642} \BibitemShut
  {NoStop}%
\bibitem [{\citenamefont {Taballione}\ \emph {et~al.}(2019)\citenamefont
  {Taballione}, \citenamefont {Wolterink}, \citenamefont {Lugani},
  \citenamefont {Eckstein}, \citenamefont {Bell}, \citenamefont {Grootjans},
  \citenamefont {Visscher}, \citenamefont {Geskus}, \citenamefont {Roeloffzen},
  \citenamefont {Renema}, \citenamefont {Walmsley}, \citenamefont {Pinkse},\
  and\ \citenamefont {Boller}}]{Taballione2019}%
  \BibitemOpen
  \bibfield  {author} {\bibinfo {author} {\bibfnamefont {C.}~\bibnamefont
  {Taballione}}, \bibinfo {author} {\bibfnamefont {T.~A.~W.}\ \bibnamefont
  {Wolterink}}, \bibinfo {author} {\bibfnamefont {J.}~\bibnamefont {Lugani}},
  \bibinfo {author} {\bibfnamefont {A.}~\bibnamefont {Eckstein}}, \bibinfo
  {author} {\bibfnamefont {B.~A.}\ \bibnamefont {Bell}}, \bibinfo {author}
  {\bibfnamefont {R.}~\bibnamefont {Grootjans}}, \bibinfo {author}
  {\bibfnamefont {I.}~\bibnamefont {Visscher}}, \bibinfo {author}
  {\bibfnamefont {D.}~\bibnamefont {Geskus}}, \bibinfo {author} {\bibfnamefont
  {C.~G.~H.}\ \bibnamefont {Roeloffzen}}, \bibinfo {author} {\bibfnamefont
  {J.~J.}\ \bibnamefont {Renema}}, \bibinfo {author} {\bibfnamefont {I.~A.}\
  \bibnamefont {Walmsley}}, \bibinfo {author} {\bibfnamefont {P.~W.~H.}\
  \bibnamefont {Pinkse}},\ and\ \bibinfo {author} {\bibfnamefont {K.-J.}\
  \bibnamefont {Boller}},\ }\href {https://doi.org/10.1364/OE.27.026842}
  {\bibfield  {journal} {\bibinfo  {journal} {Opt. Express}\ }\textbf {\bibinfo
  {volume} {27}},\ \bibinfo {pages} {26842} (\bibinfo {year}
  {2019})}\BibitemShut {NoStop}%
\bibitem [{\citenamefont {Taballione}\ \emph {et~al.}(2021)\citenamefont
  {Taballione}, \citenamefont {{van der Meer}}, \citenamefont {Snijders},
  \citenamefont {Hooijschuur}, \citenamefont {Epping}, \citenamefont {{de
  Goede}}, \citenamefont {Kassenberg}, \citenamefont {Venderbosch},
  \citenamefont {Toebes}, \citenamefont {{van den Vlekkert}}, \citenamefont
  {Pinkse},\ and\ \citenamefont {Renema}}]{Taballione2021}%
  \BibitemOpen
  \bibfield  {author} {\bibinfo {author} {\bibfnamefont {C.}~\bibnamefont
  {Taballione}}, \bibinfo {author} {\bibfnamefont {R.}~\bibnamefont {{van der
  Meer}}}, \bibinfo {author} {\bibfnamefont {H.~J.}\ \bibnamefont {Snijders}},
  \bibinfo {author} {\bibfnamefont {P.}~\bibnamefont {Hooijschuur}}, \bibinfo
  {author} {\bibfnamefont {J.~P.}\ \bibnamefont {Epping}}, \bibinfo {author}
  {\bibfnamefont {M.}~\bibnamefont {{de Goede}}}, \bibinfo {author}
  {\bibfnamefont {B.}~\bibnamefont {Kassenberg}}, \bibinfo {author}
  {\bibfnamefont {P.}~\bibnamefont {Venderbosch}}, \bibinfo {author}
  {\bibfnamefont {C.}~\bibnamefont {Toebes}}, \bibinfo {author} {\bibfnamefont
  {H.}~\bibnamefont {{van den Vlekkert}}}, \bibinfo {author} {\bibfnamefont
  {P.~W.~H.}\ \bibnamefont {Pinkse}},\ and\ \bibinfo {author} {\bibfnamefont
  {J.~J.}\ \bibnamefont {Renema}},\ }\bibfield  {journal} {\bibinfo  {journal}
  {Mater. Quantum. Technol.}\ }\href {https://doi.org/10.1088/2633-4356/ac168c}
  {10.1088/2633-4356/ac168c} (\bibinfo {year} {2021})\BibitemShut {NoStop}%
\bibitem [{\citenamefont {Ursell}(1927)}]{Ursell1927}%
  \BibitemOpen
  \bibfield  {author} {\bibinfo {author} {\bibfnamefont {H.~D.}\ \bibnamefont
  {Ursell}},\ }\href {https://doi.org/10.1017/s0305004100011191} {\bibfield
  {journal} {\bibinfo  {journal} {Mathematical Proceedings of the Cambridge
  Philosophical Society}\ }\textbf {\bibinfo {volume} {23}},\ \bibinfo {pages}
  {685} (\bibinfo {year} {1927})}\BibitemShut {NoStop}%
\bibitem [{Note1()}]{Note1}%
  \BibitemOpen
  \bibinfo {note} {See Supplemental Material}\BibitemShut {NoStop}%
\bibitem [{\citenamefont {Roeloffzen}\ \emph {et~al.}(2018)\citenamefont
  {Roeloffzen}, \citenamefont {Hoekman}, \citenamefont {Klein}, \citenamefont
  {Wevers}, \citenamefont {Timens}, \citenamefont {Marchenko}, \citenamefont
  {Geskus}, \citenamefont {Dekker}, \citenamefont {Alippi}, \citenamefont
  {Grootjans}, \citenamefont {{van Rees}}, \citenamefont {Oldenbeuving},
  \citenamefont {Epping}, \citenamefont {Heideman}, \citenamefont
  {W{\"o}rhoff}, \citenamefont {Leinse}, \citenamefont {Geuzebroek},
  \citenamefont {Schreuder}, \citenamefont {{van Dijk}}, \citenamefont
  {Visscher}, \citenamefont {Taddei}, \citenamefont {Fan}, \citenamefont
  {Taballione}, \citenamefont {Liu}, \citenamefont {Marpaung}, \citenamefont
  {Zhuang}, \citenamefont {Benelajla},\ and\ \citenamefont
  {Boller}}]{Roeloffzen2018}%
  \BibitemOpen
  \bibfield  {author} {\bibinfo {author} {\bibfnamefont {C.~G.~H.}\
  \bibnamefont {Roeloffzen}}, \bibinfo {author} {\bibfnamefont
  {M.}~\bibnamefont {Hoekman}}, \bibinfo {author} {\bibfnamefont {E.~J.}\
  \bibnamefont {Klein}}, \bibinfo {author} {\bibfnamefont {L.~S.}\ \bibnamefont
  {Wevers}}, \bibinfo {author} {\bibfnamefont {R.~B.}\ \bibnamefont {Timens}},
  \bibinfo {author} {\bibfnamefont {D.}~\bibnamefont {Marchenko}}, \bibinfo
  {author} {\bibfnamefont {D.}~\bibnamefont {Geskus}}, \bibinfo {author}
  {\bibfnamefont {R.}~\bibnamefont {Dekker}}, \bibinfo {author} {\bibfnamefont
  {A.}~\bibnamefont {Alippi}}, \bibinfo {author} {\bibfnamefont
  {R.}~\bibnamefont {Grootjans}}, \bibinfo {author} {\bibfnamefont
  {A.}~\bibnamefont {{van Rees}}}, \bibinfo {author} {\bibfnamefont {R.~M.}\
  \bibnamefont {Oldenbeuving}}, \bibinfo {author} {\bibfnamefont {J.~P.}\
  \bibnamefont {Epping}}, \bibinfo {author} {\bibfnamefont {R.~G.}\
  \bibnamefont {Heideman}}, \bibinfo {author} {\bibfnamefont {K.}~\bibnamefont
  {W{\"o}rhoff}}, \bibinfo {author} {\bibfnamefont {A.}~\bibnamefont {Leinse}},
  \bibinfo {author} {\bibfnamefont {D.}~\bibnamefont {Geuzebroek}}, \bibinfo
  {author} {\bibfnamefont {E.}~\bibnamefont {Schreuder}}, \bibinfo {author}
  {\bibfnamefont {P.~W.~L.}\ \bibnamefont {{van Dijk}}}, \bibinfo {author}
  {\bibfnamefont {I.}~\bibnamefont {Visscher}}, \bibinfo {author}
  {\bibfnamefont {C.}~\bibnamefont {Taddei}}, \bibinfo {author} {\bibfnamefont
  {Y.}~\bibnamefont {Fan}}, \bibinfo {author} {\bibfnamefont {C.}~\bibnamefont
  {Taballione}}, \bibinfo {author} {\bibfnamefont {Y.}~\bibnamefont {Liu}},
  \bibinfo {author} {\bibfnamefont {D.}~\bibnamefont {Marpaung}}, \bibinfo
  {author} {\bibfnamefont {L.}~\bibnamefont {Zhuang}}, \bibinfo {author}
  {\bibfnamefont {M.}~\bibnamefont {Benelajla}},\ and\ \bibinfo {author}
  {\bibfnamefont {K.-J.}\ \bibnamefont {Boller}},\ }\href
  {https://doi.org/10.1109/JSTQE.2018.2793945} {\bibfield  {journal} {\bibinfo
  {journal} {IEEE Journal of Selected Topics in Quantum Electronics}\ }\textbf
  {\bibinfo {volume} {24}},\ \bibinfo {pages} {1} (\bibinfo {year}
  {2018})}\BibitemShut {NoStop}%
\bibitem [{\citenamefont {Clements}\ \emph {et~al.}(2016)\citenamefont
  {Clements}, \citenamefont {Humphreys}, \citenamefont {Metcalf}, \citenamefont
  {Kolthammer},\ and\ \citenamefont {Walmsley}}]{Clements2016}%
  \BibitemOpen
  \bibfield  {author} {\bibinfo {author} {\bibfnamefont {W.~R.}\ \bibnamefont
  {Clements}}, \bibinfo {author} {\bibfnamefont {P.~C.}\ \bibnamefont
  {Humphreys}}, \bibinfo {author} {\bibfnamefont {B.~J.}\ \bibnamefont
  {Metcalf}}, \bibinfo {author} {\bibfnamefont {W.~S.}\ \bibnamefont
  {Kolthammer}},\ and\ \bibinfo {author} {\bibfnamefont {I.~A.}\ \bibnamefont
  {Walmsley}},\ }\href {https://doi.org/10.1364/OPTICA.3.001460} {\bibfield
  {journal} {\bibinfo  {journal} {Optica}\ }\textbf {\bibinfo {volume} {3}},\
  \bibinfo {pages} {1460} (\bibinfo {year} {2016})}\BibitemShut {NoStop}%
\bibitem [{\citenamefont {Evans}\ \emph {et~al.}(2010)\citenamefont {Evans},
  \citenamefont {Bennink}, \citenamefont {Grice}, \citenamefont {Humble},\ and\
  \citenamefont {Schaake}}]{Evans2010}%
  \BibitemOpen
  \bibfield  {author} {\bibinfo {author} {\bibfnamefont {P.~G.}\ \bibnamefont
  {Evans}}, \bibinfo {author} {\bibfnamefont {R.~S.}\ \bibnamefont {Bennink}},
  \bibinfo {author} {\bibfnamefont {W.~P.}\ \bibnamefont {Grice}}, \bibinfo
  {author} {\bibfnamefont {T.~S.}\ \bibnamefont {Humble}},\ and\ \bibinfo
  {author} {\bibfnamefont {J.}~\bibnamefont {Schaake}},\ }\href
  {https://doi.org/10.1103/PhysRevLett.105.253601} {\bibfield  {journal}
  {\bibinfo  {journal} {Phys. Rev. Lett.}\ }\textbf {\bibinfo {volume} {105}},\
  \bibinfo {pages} {253601} (\bibinfo {year} {2010})}\BibitemShut {NoStop}%
\bibitem [{\citenamefont {Marsili}\ \emph {et~al.}(2011)\citenamefont
  {Marsili}, \citenamefont {Najafi}, \citenamefont {Dauler}, \citenamefont
  {Bellei}, \citenamefont {Hu}, \citenamefont {Csete}, \citenamefont {Molnar},\
  and\ \citenamefont {Berggren}}]{Marsili2011}%
  \BibitemOpen
  \bibfield  {author} {\bibinfo {author} {\bibfnamefont {F.}~\bibnamefont
  {Marsili}}, \bibinfo {author} {\bibfnamefont {F.}~\bibnamefont {Najafi}},
  \bibinfo {author} {\bibfnamefont {E.}~\bibnamefont {Dauler}}, \bibinfo
  {author} {\bibfnamefont {F.}~\bibnamefont {Bellei}}, \bibinfo {author}
  {\bibfnamefont {X.}~\bibnamefont {Hu}}, \bibinfo {author} {\bibfnamefont
  {M.}~\bibnamefont {Csete}}, \bibinfo {author} {\bibfnamefont {R.~J.}\
  \bibnamefont {Molnar}},\ and\ \bibinfo {author} {\bibfnamefont {K.~K.}\
  \bibnamefont {Berggren}},\ }\href {https://doi.org/10.1021/nl2005143}
  {\bibfield  {journal} {\bibinfo  {journal} {Nano Lett.}\ }\textbf {\bibinfo
  {volume} {11}},\ \bibinfo {pages} {2048} (\bibinfo {year}
  {2011})}\BibitemShut {NoStop}%
\bibitem [{\citenamefont {Holzman}\ and\ \citenamefont
  {Ivry}(2019)}]{Holzman2019}%
  \BibitemOpen
  \bibfield  {author} {\bibinfo {author} {\bibfnamefont {I.}~\bibnamefont
  {Holzman}}\ and\ \bibinfo {author} {\bibfnamefont {Y.}~\bibnamefont {Ivry}},\
  }\href {https://doi.org/10.1002/qute.201800058} {\bibfield  {journal}
  {\bibinfo  {journal} {Advanced Quantum Technologies}\ }\textbf {\bibinfo
  {volume} {2}},\ \bibinfo {pages} {1800058} (\bibinfo {year}
  {2019})}\BibitemShut {NoStop}%
\bibitem [{\citenamefont {Feito}\ \emph {et~al.}(2009)\citenamefont {Feito},
  \citenamefont {Lundeen}, \citenamefont {{Coldenstrodt-Ronge}}, \citenamefont
  {Eisert}, \citenamefont {Plenio},\ and\ \citenamefont
  {Walmsley}}]{Feito2009}%
  \BibitemOpen
  \bibfield  {author} {\bibinfo {author} {\bibfnamefont {A.}~\bibnamefont
  {Feito}}, \bibinfo {author} {\bibfnamefont {J.~S.}\ \bibnamefont {Lundeen}},
  \bibinfo {author} {\bibfnamefont {H.}~\bibnamefont {{Coldenstrodt-Ronge}}},
  \bibinfo {author} {\bibfnamefont {J.}~\bibnamefont {Eisert}}, \bibinfo
  {author} {\bibfnamefont {M.~B.}\ \bibnamefont {Plenio}},\ and\ \bibinfo
  {author} {\bibfnamefont {I.~A.}\ \bibnamefont {Walmsley}},\ }\href
  {https://doi.org/10.1088/1367-2630/11/9/093038} {\bibfield  {journal}
  {\bibinfo  {journal} {New J. Phys.}\ }\textbf {\bibinfo {volume} {11}},\
  \bibinfo {pages} {093038} (\bibinfo {year} {2009})}\BibitemShut {NoStop}%
\bibitem [{Note2()}]{Note2}%
  \BibitemOpen
  \bibinfo {note} {During data analysis, it became apparent that due to a
  communication error, that the second phase shifter of the top-left unit cell
  was not actuated on during measurements. This does not significantly affect
  the results of the Haar measurements.}\BibitemShut {Stop}%
\end{thebibliography}
\end{document}